\documentclass{article}

\usepackage{arxiv}
\usepackage[utf8]{inputenc}
\usepackage[T1]{fontenc}
\usepackage{hyperref}
\usepackage{url}
\usepackage{booktabs}
\usepackage{amsfonts}
\usepackage{nicefrac}
\usepackage{microtype}
\usepackage{graphicx}
\usepackage{csquotes}
\usepackage{epstopdf,epsfig}
\usepackage{newtxtext}
\usepackage{newtxmath}
\usepackage{subfigure}
\usepackage{natbib}
\usepackage{xcolor}
\usepackage{tikz}
\hypersetup{
	colorlinks = true,
	urlcolor   = blue,
	citecolor  = black,
}

\definecolor{lightgreen}{rgb}{0.56, 0.93, 0.56}  
\definecolor{lightblue}{rgb}{0.68, 0.85, 0.90}   

\title{The influence of free-stream turbulence on the fluctuating loads experienced by a cylinder exposed to a turbulent cross-flow}

\author{Francisco J. G. de Oliveira \\
        Department of Aeronautics\\
        Imperial College London, London, UK\\
        \texttt{f.oliveira22@imperial.ac.uk}\\
        \And
        Zahra Sharif Khodaei\\
        Department of Aeronautics\\
        Imperial College London, London, UK\\
        \texttt{z.sharif-khodaei@imperial.ac.uk}\\
        \And
        Oliver R. H. Buxton\\
        Department of Aeronautics\\
        Imperial College London, London, UK\\
        \texttt{o.buxton@imperial.ac.uk}\\
        }

\begin{document}
\maketitle
\begin{abstract}
	The impact of several \enquote{flavours} of free-stream turbulence (FST) on the structural response of a cantilevered cylinder, subjected to a turbulent cross-flow is investigated. At high enough Reynolds numbers, the cylinder generates a spectrally rich turbulent wake which significantly contributing to the experienced loads. The presence of FST introduces additional complexity through two primary mechanisms: \textbf{directly}, by imposing a fluctuating velocity field on the cylinder's surface, and \textbf{indirectly}, by altering the vortex shedding dynamics, modifying the experienced loads. We employ concurrent temporally resolved Particle Image Velocimetry (PIV) and distributed strain measurements using Rayleigh backscattering fibre optic sensors (RBS) to instrument the surrounding velocity field and the structural strain respectively. By using various turbulence-generating grids, and manipulating their distance to the cylinder, we assess a broad FST parameter space allowing us to individually explore the influence of transverse integral length scale ($\mathcal{L}_{13}/D$), and turbulence intensity ($TI$) of the FST on the developing load dynamics. The presence of FST enhances the magnitude of the loads acting on the cylinder. This results from a decreased vortex formation length, increased coherence of regular vortex shedding, and energy associated with this flow structure in the near-wake. The cylinder's structural response is mainly driven by the vortex shedding dynamics, and their modification induced by the presence of FST, ie. the indirect effect outweighs the direct effect. From the explored FST parameter space, $TI$ was seen to be the main driver of enhanced loading conditions, presenting a positive correlation with the fluctuating loads magnitude at the root. 
\end{abstract}


\section{Introduction}\label{sec:Introduction}

Free-stream turbulence (FST) plays a pivotal role in a broad range of engineering applications, influencing the lifetime and performance of structures in several situations. Fuel elements in nuclear reactors \citep{wang2019}, wind turbines in wind farms \citep{chamorro2011, stevens2017} and buildings in cities \citep{hiromasa2011} are often designed to accommodate the effects of FST. The widespread presence of FST in natural flows also influences how trees grow in forests, and is relevant to predict forest damage induced by climate change and the modification of wind conditions \citep{sellier2008}. The structures above share a common feature: they act as bluff bodies. When exposed to a cross-flow, an unsteady flow is generated, exposing themselves to the phenomenology of the developing flow. The study of the flow around bluff bodies holds paramount significance in fluid mechanics, spanning several decades.

The cross-flow over a circular cylinder, a quintessential bluff body, has been a focal point of intensive research spanning numerous works due to its duality in the simplicity and complexity of the occurring flow phenomena. Pioneering contributions by \cite{roshko1954} and \cite{gerrard1966} elucidated the fundamental aspects of vortex formation, wake characteristics and flow-generated loads on this body. The comprehensive reviews of \cite{bearman1983} and \cite{williamson1996} further distil this knowledge. The Reynolds number plays a pivotal role in the development of the flow conditions. After a specific regime, the interplay between a cylinder's free shear layers gives rise to regular vortex shedding. This phenomenon ends up dominating the loading events experienced by the cylinder exposed to the cross-flow. Regular vortex shedding engenders a mean drag force and a temporal oscillation of lift and drag, generated by the oscillating pressure field developing around the cylinder.

A general case of the flow around a circular cylinder is the flow around finite cylinders \citep{porteous2014, crane2021, okamoto1973, farivar1981, fox1993a, fox1993b, fox1993c, essell2021, uematsu1990effects}. Besides the dependence on the Reynolds number ($Re$), the flow is highly sensitive to the cylinder's aspect ratio $L/D$. The three-dimensionality of this flow is frequently met in several engineering applications, such as wind turbine masts or building wakes \citep{Wangandfan2019,Wang2018}, where the $3$D free end effects generate a more complex shedding process than the classical $2$D flow case \citep{crane2021}. \cite{porteous2014} provided a comprehensive review of several models to describe the wake topology of such flows, and \cite{essell2021} analysed how different aspect ratios influence the shedding from the cylinder. Furthermore, \cite{fox1993a, fox1993b, fox1993c} explored the effects of different aspect ratios on the loads experienced by the cylinder, analysing the sensitivity of the flow to the aspect ratio. In addition, \cite{okamoto1973} verified that the introduced vorticity and entrainment near the wake by $3D$ flow events acting on the wake relieves some low pressure in the wake, decreasing the drag of the cylinder when compared to an infinite cylinder. Overall, this generates an inhomogeneous shedding of the von K\'{a}rm\'{a}n street, where the intensity of this flow structure grows towards the root, imposing obliqueness into the vortex shedding. Depending on the aspect ratio of the cylinder, the shedding process of the cylinder changes severely. In our experiment, the cylinder is set with an aspect ratio of $L/D \sim 10$. In these conditions, two cellular shedding regions along the cylinder, where close to the free end, longitudinal streamwise vortices develop. Away from these and closer to the root, the shedding process is characterised by regular vortex shedding \citep{porteous2014}. 

The presence of FST has been associated with a modification of the entrainment of momentum, energy and mass into the near, and far wake of a circular cylinder \citep{jiangang2023, krishna2020, krishna2023}. The increase of the integral length scale ($\mathcal{L}$) in the turbulent free-stream past a cylinder has been associated with promotion of large-scale engulfment in the near wake. Contrastingly, the increase of turbulence intensity ($TI$) in the free-stream has been associated with a reduced entrainment rate in the far wake of a cylinder \citep{krishna2020}, and with an increase of vibrations, and forces experienced by bluff bodies \citep{bearman1983, wang2019}. This is in line with the reported decrease of the recirculation region and extent of the vortex formation region, pulling low-pressure vortex cores closer to the cylinder \citep{norberg1987, west1993}. Furthermore, the presence of FST accelerates the transition to turbulence of the attached boundary layers, modifying the position of the flow separation points over the surface of the cylinder \citep{bearman1983}, which in turn modifies the wake expansion and evolution when compared to cases where no FST is present. \cite{maryami2019} showed how FST influences the aerodynamic performance of a double fixed cylinder. Combining the usage of sets of turbulence-generating grids, and pressure transducers over the cylinder's spanwise extent, the authors reported an increased extent of spanwise correlation at the regular vortex shedding frequency with an increase of $TI$. 

FST has been associated with an increased vibration, and larger fluctuating loads on bluff bodies under cross-flows \citep{wang2019, francisco2024,bearman1983,uematsu1990effects,ramesh2024vortex}. We may separate the impact of FST on the loads acting over a wake-generating body into two effects: a direct impact by introducing an increasingly energetic fluctuating velocity field into the free-stream, immediately impacting on the windward surface of the body; and an indirect impact, modifying the regular shedding process within the cylinder's wake thereby impacting on the dynamic load experienced by it. \cite{francisco2024} analysed the response of a cantilevered cylinder exposed to free-stream turbulence, and in this work, the authors identified a stronger correlation of the structural dynamics to the modification of vortex shedding when compared to a case without the presence of FST, being this flow structure the most relevant to the structural response of the cylinder. \cite{ramesh2024vortex} also reported an increased energy of regular vortex shedding with the introduction of moderate levels of FST. We explore in depth the impact introduced by different FST \enquote{flavours}, isolating the impact of both $\mathcal{L}/D$ and $TI$, and the different dynamics imposed which then correlate flow to the structural dynamics. 

To measure flow-induced loads, load cells, strain gauges, microphones, pressure taps or pressure transducers are often employed. Load cells are usually connected to the base of bodies, so are thus restricted to measuring the integrated effect of flow events occurring over the entirety of the body, and pressure taps and pressure transducers as used by \cite{maryami2019} allow the acquisition of spatially distributed information, at the expense of complicated and burdensome setups. To obtain the strain field of submerged structures (in water), strain gauges are deemed unusable unless using waterproof coatings, which end up affecting the aerodynamic profile of the body. Fibre optic sensors \citep{francisco2024, zhou1999, zhou2000, Xu2020} have the potential to overcome this problem. Their dimensions and small weight have been proven to have a negligible effect on the structural response, and developing flow field \citep{zhou1999}. In the current experimental campaign, we employ novel fibre optic Rayleigh backscattering sensors (RBS) \citep{francisco2024, Xu2020}, to obtain the structural response of a cantilevered cylinder to different free-stream and flow conditions, along the spanwise direction of the structure. Fibre optic sensors such have been previously used to instrument flow-induced strain over submerged structures \citep{zhou1999, zhou2000, wang2019, delatorre2021, zhou2001}. This type of sensor has shown its compatibility to accurately represent the bending displacement on a structure induced by a cross-flow, with and without the presence of FST, and within the neighbouring regions of other bluff bodies \citep{zhou2001}. RBS rely on an optical frequency-domain reflectometry (OFDR) system, which exploits the Rayleigh backscattering spectrum occurring in the single mode fibres (SMF) generated by random modulation of the refractive index profile along the length of the fibre yielded by the local strain field, to instrument and capture details of the structural response of the analysed body. The modulation of the refractive index within different positions of the fibre changes according to the response of the structure to which the fibres are integrated, allowing the assessment of the structural behaviour of different bodies subjected to the different dynamic conditions \citep{Xu2020}. These measurements can be directly correlated to local strain/stresses experienced by the structure. In addition, thanks to the high spatial resolution and large extent of the fibres, we are able to easily instrument $520$ sensing points with a resolution of $2.6$mm using a single channel. This strain can then be directly correlated to the local loads acting on the structure.

The present investigation sheds light on how different FST conditions modify the developing flow around a cantilevered cylinder with an aspect ratio of $L/D \sim 10$, and how these modifications impact on experienced loads. This is done by combining a RBS network and $2$D PIV simultaneously, allowing us to spatially and temporally resolve the strain, and flow field surrounding the cylinder. This novel technique was developed in \cite{francisco2024}. By doing so, we are able to assess different flow conditions and their respective impact on the structural response of the cylinder. The $2$D-PIV captures the flow in two transverse planes to analyse how FST affects the shedding events of the cylinder, and to analyse how the flow structures captured in these planes map to the structural response of the body. The influence of FST on the flow, and structural response of the cylinder was analysed by exploring the turbulence intensity $TI$ and integral length scale $\mathcal{L}_{13}/D$ space upstream of the cylinder. Furthermore, the used fibre optic strain sensing network allows us to distinguish the strain experienced locally by the cylinder upstream and downstream of the position of the time-averaged shear-layer separation, enabling us to differentiate the effects of the modification of the near wake (by the FST), from the direct impact of the FST on the cylinder.

Following this introduction, section \S\ref{sec:Methodology} details the experimental apparatus and methodology used to obtain the current set of data. Major obtained results and conclusions are presented throughout sections \S \ref{sec:results1}-\S\ref{sec:results4}. Section \S \ref{sec:Conclusions} presents a summary of the outcome of this experimental campaign.

\section{Experimental Methodology}\label{sec:Methodology}

\begin{figure}
	\centering
	\hspace*{-0.3cm}{
		\includegraphics[width = 0.7\textwidth]{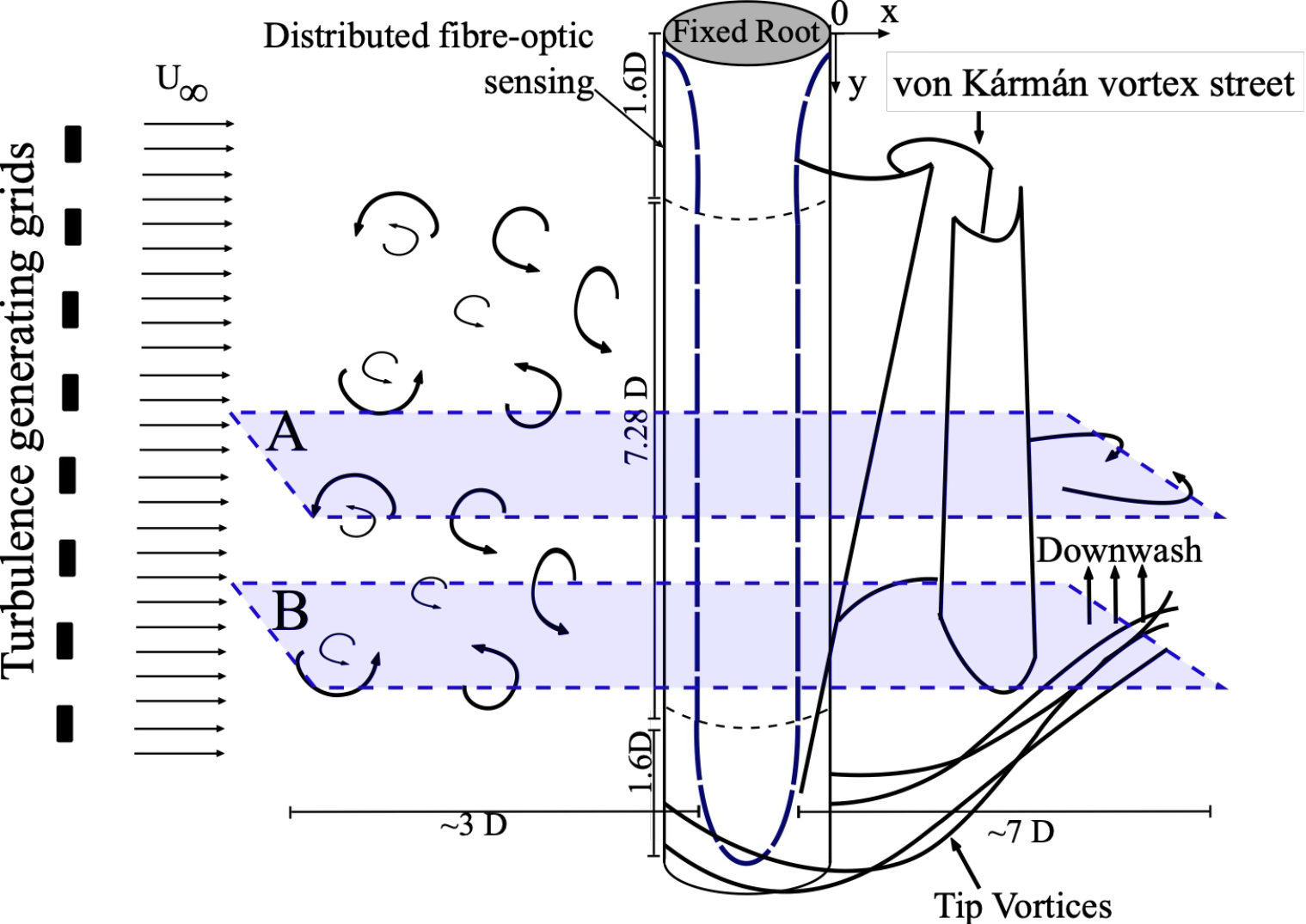}\hspace*{-0.cm}}
	\caption{Experimental schematic layout: (\textit{a})- fibre optic sensing path, fields of view (FOV) captured and representation of the main flow events over the cylinder. The used Cartesian space of coordinates is represented in the figure, where $y$ corresponds to the spanwise direction of the cylinder, $x$ to the streamwise and $z$ to the transverse direction of the flow.}
	\label{fig:exp_set_up}
\end{figure}

\begin{table}
	\centering{
		\begin{tabular}{cccccc}
			FOV          & Spanwise position & FOV dimensions                                                                   & $\Delta x$ & $N_{vs}$  \\
			A & $ y = 3.4 D $     & \begin{tabular}[c]{@{}c@{}}$-5.7 < x /D< 7.4$,\\ $-2.5 < z/D < 2.2$\end{tabular} & $0.024D$   & $\approx 250$ \\
			B& $ y = 7.4 D $     & \begin{tabular}[c]{@{}c@{}}$-5.1 < x /D< 7.4$,\\ $-2.2 < z/D < 1.8$\end{tabular} & $0.020D$  & $\approx 250$ 
	\end{tabular}}
	\caption{Experimental conditions for the different captured FOVs. $\Delta x$ and $N_{vs}$ correspond, respectively, to the spatial resolution of the experiments for each acquired FOV, and to the number of vortex-shedding cycles captured.}
	\label{tab:exp_conditions}
	\vspace{0.5cm}
	\centering{
		\begin{tabular}{cccccccccc}
			Case             & $1a$ & $1b$ & $1c$ & $2a$ & $2b$ & $2c$ & $2d$ & $3a$ & $3b$ \\ 
			$U_{1}$ [m/s] & $0.46$ & $0.52$ & $0.52$ & $0.60$ & $0.53$ & $0.51$ & $0.49$ & $0.56$ & $0.53$
		\end{tabular}}
	\caption{Free-stream incoming flow velocities for each FST case.}
	\label{tab:Uinf_fstcases}
\end{table}

\begin{figure}
	\centering
	\hspace*{-0.3cm}{
		\raisebox{2.1in}{\textit{a})}\raisebox{0.in}{\includegraphics[width = 0.47\textwidth]{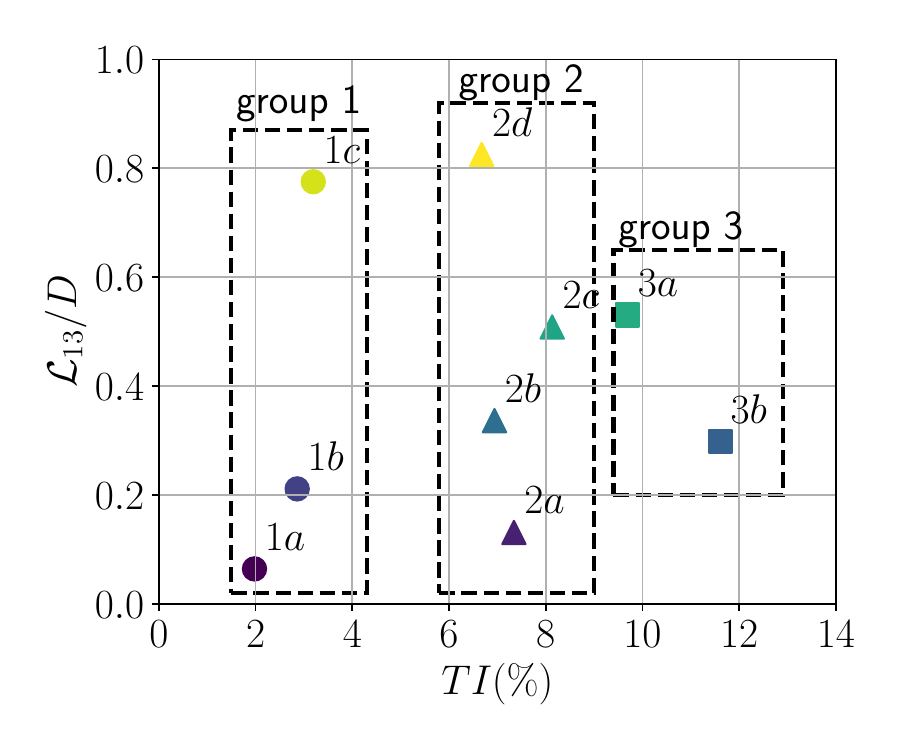}}\hspace*{-0.cm}}
	\raisebox{2.1in}{\textit{b})}\includegraphics[width = 0.53\textwidth]{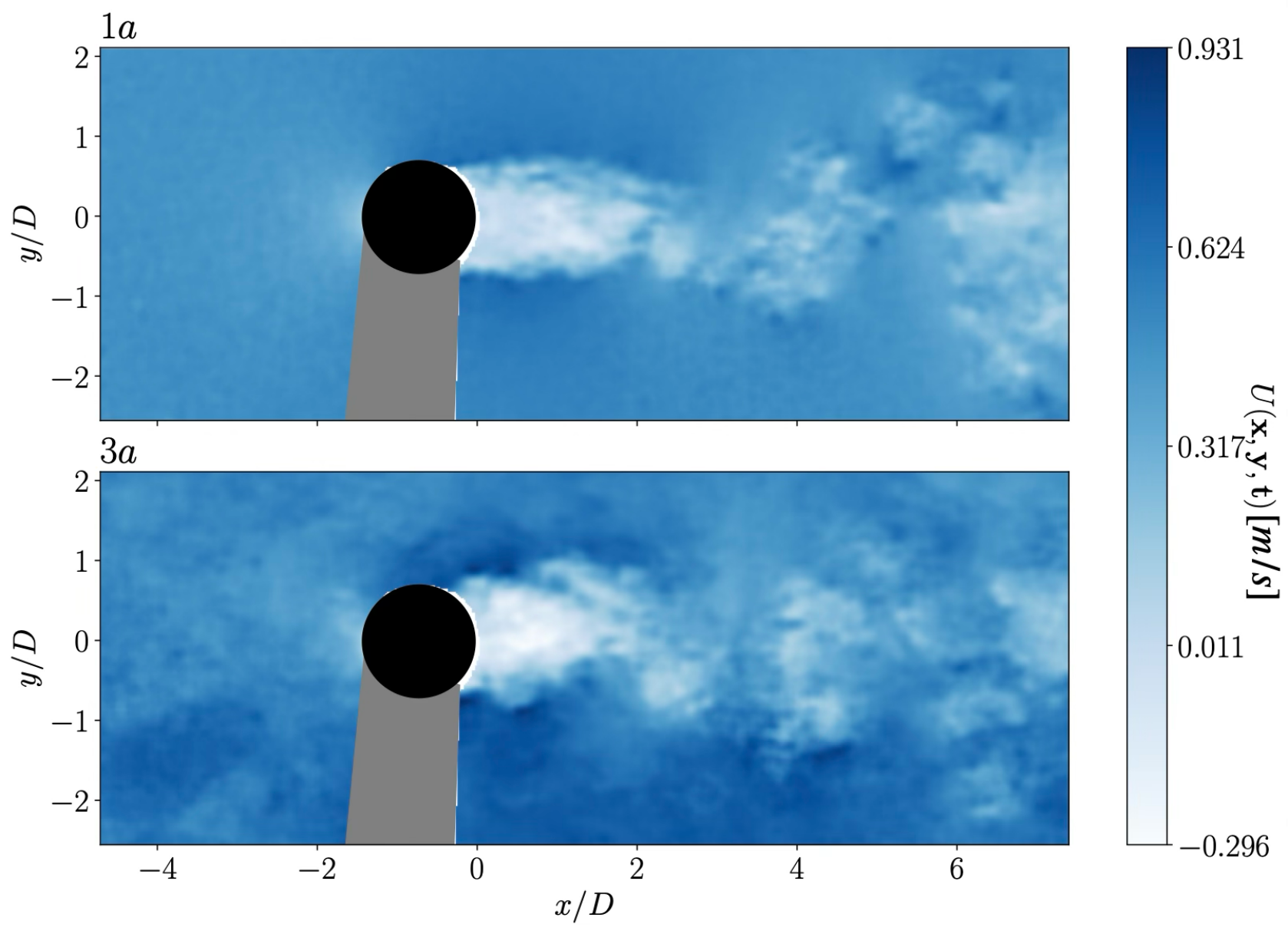}\hspace*{-0.cm}
	\caption{ (\textit{a}): FST $\{\mathcal{L}_{13}/D , TI\%\}$ parameter space tested. Groups are defined based on their $TI$, to explore within each group the effect of increasing $\mathcal{L}_{13}$. (\textit{b}): PIV snapshots captured in FOV A for the FST conditions described by cases $1a$, no FST, and $3a$, high $TI$ and medium $\mathcal{L}_{13}$. The black circle corresponds to the position of the cylinder during the experimental tests, and the grey region to the shadow originating from the positioning of the cylinder with respect to the laser sheet.}
	\label{fig:FSTcond}
\end{figure}

The experimental campaign was conducted in the hydrodynamics flume of the Department of Aeronautics at Imperial College London. A novel combination of concurrent measurements of RBS sensors and PIV was employed. Readers are referred to \cite{francisco2024} for more details on the used methodology. The water tunnel was kept with a constant cross-section of $58\times60$ cm$^2$ during the experiments. The test specimen consisted of an acrylic cylinder with $D = 50$ mm diameter, and $2.5$ mm wall thickness ($l_t$). The cylinder was submerged with $95\%$ of its body underwater, representing a wet length of $\approx 500$ mm (see figure \ref{fig:exp_set_up})). During the submersion of the cylinder, the cylinder was also filled with water to eliminate buoyancy effects. The tip of the cylinder was kept at a distance $80$ mm away from the bottom of the tunnel, allowing the development of tip vortices and their respective wake advection towards the fixing root of the cylinder. The distance to the bottom of the flume also allowed the bottom wall boundary layer to develop without interacting with the cylinder. The cylinder was fixed as a cantilever, restraining any rotation and movement at the root. The incoming velocity without the presence of the turbulence-generating grids was $U_{1a} = 0.46$ m/s, setting the Reynolds number based on the cylinder's diameter to $Re \approx 23,000$. Surface deformation is not expected to interact with the developing flow dynamics over the cylinder as the computed Froude number was kept to $Fr \approx U_1/\sqrt{g\times H} \approx 0.2$ (\cite{cicolin2021}). Two PIV fields of view were captured for each of the tested FST conditions, with concurrent strain data.  In each run, the flow was initiated by gradually increasing the speed of the flume until the desired flow velocity was achieved. Data acquisition was triggered once the flow had reached a stable state. The Strouhal number is henceforth defined as $St = f \times D/U_1$, where $U_1$ corresponds to the incoming mean velocity under each experimental FST condition. This definition takes into account the dynamic modification introduced by an increase in the incoming mean velocity field, caused by the pressure drop introduced by the presence of the turbulence generating grid which yields a small decrement of the water depth. The experienced free-stream velocities for each of the tested cases are presented in table \ref{tab:Uinf_fstcases}. The averaged experienced Reynolds number by the cylinder was $Re \approx 26,000$. The maximum free-stream velocity experienced by the cylinder was $U_{2b} = 0.6$ m/s. The cylinder's drag regime thus did not change throughout the test cases reported. The used Cartesian system is centred at the top of the cylinder, on the rear face of its fixing region (see figure \ref{fig:exp_set_up}), and the fluctuating velocity components obtained by applying a Reynolds decomposition to the flow velocity corresponding to $\{x,y,z\}$ will herein be referred to $u_i$, where $i \in [1,2,3]$.

Following \cite{jiangang2023, krishna2020, krishna2023}, we change the FST conditions to which the cylinder is exposed by manipulating the turbulence intensity ($TI = \sqrt{(u_1^2+u_3^2)/2}/U_1$), and integral length scale ($\mathcal{L}_{13}/D = \int_0^{r_i} R_{13}(r)\mathrm{d}r/D$) of the incoming flow. The cylinder was then exposed to different \enquote{flavours} of FST, by exploring the FST parameter space of $\{TI, \mathcal{L}_{13}/D\}$. In the previous definitions, $u_i$ corresponds to the velocity fluctuations, $R_{13}$ corresponds to the auto-correlation function between $u_{1}(x,z)$ and $u_{1}(x,z+r)$, and $r_i$ to the first zero crossing position of the auto-correlation function. These statistics were computed $2D$ upstream of the cylinder. 
 
Figure \ref{fig:FSTcond} \textit{a)} details the $9$ \enquote{flavours} of FST considered. Table \ref{tab:Uinf_fstcases} presents the averaged free-stream velocities for each of the explored cases represented. The test cases were grouped with respect to their $TI$ levels, representing low, moderate, and high $TI$ conditions - respectively groups $[1,2,3]$. This was done to individualise within each group of similar $TI$, the relative influence of $\mathcal{L}_{13}/D$ on the experienced loads by the cylinder. Figure \ref{fig:FSTcond} \textit{b)} corresponds to two instantaneous snapshots captured by FOV A, for cases $1a$ and $3a$. Due to space constraints, the largest integral length scale achieved in the free-stream of the cylinder is smaller than $1D$. Larger integral length scales were not tested to preserve uniformity of the temporally averaged flow upstream of the cylinder. This ensured a maximum gradient of $10\%$ of the incoming temporally-averaged velocity fields, across the transverse plane of the cylinder. For the remainder of the manuscript, $U_i$ refers to the velocity field as captured in the PIV FOV, and $u_i$ to the fluctuating velocity field after applying a Reynolds decomposition, removing the time-averaged velocity $\overline{U_{i}}$ from $U_i$. In addition to generated free-stream turbulence, the presence of the turbulence-generating grids accounts for an increase of the incoming mean velocity, caused by the increased blockage introduced by the presence of an obstructing element to the flow.

\subsection{Temporally resolved $2$D-PIV} \label{sec:2dpivexpmeth}

The analysis of the flow developing around the cylinder was achieved by performing $2$D-PIV in two different planes, at two different spanwise locations. This was done to document how the different spanwise cellular shedding mechanisms influenced the structural response, and how they were affected by the presence of FST. Field of View (FOV) A was located at $y/D = 3.4$, away from the $3$D flow effects developed at the free end of the cylinder characterising the regular vortex-shedding cell at the midspan \citep{porteous2014}; and FOV B at $y/D = 7.4$, characterising the longitudinal tip vortices shed from the free end of the cylinder. Table \ref{tab:exp_conditions} outlines the experimental details on the resolution and acquisition of the PIV. For both FOVs, the streamwise and transverse extent was similar. Both FOVs documented the region upstream and downstream of the cylinder.  

The illumination source for the PIV measurements was a high-speed Litron LDY304 Nd:YLF laser, operating at $1000$ Hz. The images were acquired with two Phantom V640L cameras. The resulting FOVs resulted from the combination of the two velocity fields acquired, mapping the velocity field into the same coordinate referential. The cameras acquired a total of $12372$ images each, at full resolution ($2560$ px $\times 1600$ px), with an acquisition frequency of $f_{\text{acq}} = 50$ Hz. This resulted in $\approx 125$ s of data acquisition, $\approx 250$ vortex shedding cycles. The cameras were connected to two Nikkor lenses of $50$ mm, set with an aperture of $f^{\#} = 2.8$.  The resulting images were processed employing a multi-pass PIV algorithm, to estimate the average particle displacement per interrogation window. $4$ passes were used, starting with a $32\times 32$ window, ending up with a $16\times 16$ window, with a $50\%$ overlap region in each interrogation step. The spurious vector field was identified with a local $5\times 5$ mean test. The number of spurious vectors in the used snapshot groups was less than $1\%$, and each vector, after identification, was recomputed by a linear interpolation of the neighbouring vector field.

\subsection{Concurrent RBS measurements}

The RBS sensors were integrated onto the cylinder's surface, as represented in figure \ref{fig:exp_set_up}, to accurately interrogate the structural response of the cylinder subjected to the different FST \enquote{flavours}. The fibre-optic sensing range was entirely submerged, and the polar positions of the fibres were set using an optical rotating table. The strain data was acquired with the OFDR system developed by LUNA Ltd - ODiSI-B. Single mode fibre optics (SMF) were used as RBS in this study, to interrogate the cylinder's strain response, along its spanwise direction.  The sensing network configuration enables the assessment of the direct impact of FST (on the windward face of the cylinder), through consideration of the distinct interactions between the wake and free-stream conditions, and the leeward ($\theta_f^{\pm 135^{\circ}}$) and windward ($\theta_f^{\pm 45^{\circ}}$) fibres attached to the cylinder \citep{francisco2024}.

The strain data was acquired at a frequency of $50$ Hz, and filtered with a low-pass filter of $15$ Hz. The low-pass frequency was set well above the characteristic frequency of the largest shed structures by the cylinder ($2$Hz), to remove high-frequency noise from the strain signal produced by the harmonics of the structure. An analysis of the PIV data shows that this operation does not remove the influence of vortex shedding and other important flow strucutres from the fibre optic signal. Each sensing line over each polar coordinate acquired a total of $140$ sensing points, with a spatial resolution of $\Delta y = 2.6$ mm ($\Delta y/D \approx 0.05$). The instrumented region spanned for an extent of $L_{i}=7.28D$ with an offset of $y/D = 1.6$ both at the root and tip, to allow enough room to pass the fibres, without compromising the maximum bending radius \citep{francisco2024}. For each FST condition, the strain data was recorded twice, concurrently with PIV acquisitions in both FOVs. For more details on the methodology of the data acquisition, consult \cite{francisco2024}. As the cylinder's length is larger than the extent of the instrumented region, we define herein a local referential coordinate $y'$, spanning from the start of the measurement region, and until its end at $y/L_{i}=1$.

\section{FST influence on the time-averaged loads of the cylinder}\label{sec:results1}

The presence of FST has been previously associated with a decrease of the vortex formation length \citep{khabouchi2014,bearman1983,gerrard1966} on the wake of a cylinder. This in turn, modifies the mean-flow around the cylinder which results in a modification of the cylinder's experienced time-averaged drag \cite{bearman1983, uematsu1990effects}. To explore the effect of the tested FST parameter space on the cylinder's time-averaged loads, we analyse the cylinder's time-averaged tip deflection. This metric reflects the overall induced force on the body, which is directly related to its maximum deflection magnitude. Thanks to the large sensing density provided by the RBS, we are able to reconstruct the deflection field of the cylinder from the measured strain, with the following shape sensing algorithm \citep{Xu2020}:

\begin{equation}	
	\begin{split}
		\beta_n = \frac{2\Delta L \times \varepsilon_{n-1}}{l_t} + \sum_{j=1}^{n-1}\beta_j  \qquad \qquad \text{(i)}\\
		\delta_{n} = \Delta L \times \left(\frac{2\Delta L\times\varepsilon_{n-1}}{l_t} + \beta_{n-1}\right) + \delta_{n-1}, \qquad \text{(ii)}
	\end{split}
	\label{eq:shape_sensing_algorithm}
\end{equation}
where $\beta_n$ and $\delta_n$ correspond respectively to the rotation and deflection of the local sensorised regions. The deflection of the body is reconstructed using the local deformation obtained by the strain sensors, by projecting the acquired strain together with the local rotation experienced by each node (see equation \eqref{eq:shape_sensing_algorithm}). To compute the deflection field, we assume negligible deflection before the start of the sensing network ($\delta(y/L{i} = 0)=0$). We then define the cylinder's tip deflection as:
\begin{equation}
	\delta_{tip} = \frac{1}{N_{fibres}}\sum_i^{N_{fibres}}\overline{\delta(t,i,y'/L_{i} = 1)},
\end{equation}
reflecting the ensemble average of tip deflection captured by the 4 sets of fibres used, at $y'/L_{i} = 1$.

\begin{figure}
    \centering
    \includegraphics[width=\textwidth]{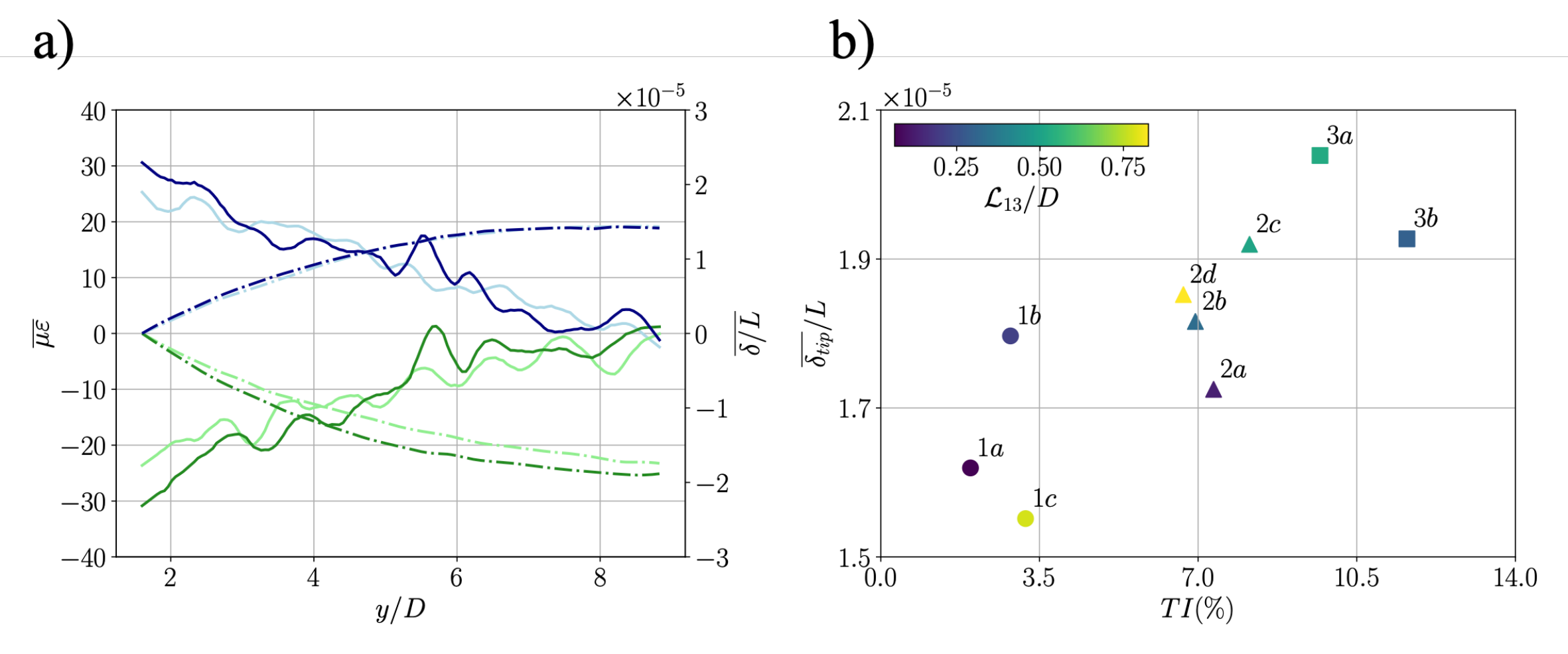}
    \caption{(\textit{a}): Time-averaged strain ($\overline{\varepsilon}$)(\raisebox{0.5ex}{\textcolor[HTML]{000000}{\rule{0.25cm}{0.9pt}}}) and reconstructed deflection field ($\overline{\delta}$) (\raisebox{0.5ex}{\textcolor[HTML]{000000}{\rule{0.125cm}{0.9pt}}}  \textcolor[HTML]{000000}{\rule{0.05cm}{0.9pt}}) obtained by $\theta_f^{\alpha}$, $\alpha \in [45, -45, 135, -135]^{\circ}$ respectively represented by \raisebox{0.5ex}{\textcolor[HTML]{00008B}{\rule{0.25cm}{0.9pt}}}, \raisebox{0.5ex}{\textcolor[HTML]{ADD8E6}{\rule{0.25cm}{0.9pt}}}, \raisebox{0.5ex}{\textcolor[HTML]{006400}{\rule{0.25cm}{0.9pt}}} and \raisebox{0.5ex}{\textcolor[HTML]{90EE90}{\rule{0.25cm}{0.9pt}}}. The waviness in the strain distribution is related to the non-uniformity wall thickness of the cylinder, due to its manufacturing process. (\textit{b}): Averaged normalised tip deflection ($\delta_{\text{tip}}/L$) for each FST case. }.
    \label{fig:Figure_3.1}
\end{figure}

Figure \ref{fig:Figure_3.1} \textit{(a)} corresponds to the captured time-averaged strain field, and reconstructed deflection field from the set of $4$ fibres, for case $1a$. Figure \ref{fig:Figure_3.1} \textit{a)} shows an increased magnitude of strain at the root of the cylinder, negative for the leeward face of the cylinder representative of the compressive loads in this region of the cylinder, and positive on the windward face of the cylinder, representative of tensile loads. The root region consists of the region with the largest magnitude of instrumented strain, whilst presenting negligible deflections. For a cantilevered structure this is the location of the peak bending moment, and the region where these bodies are prone to fail. In contrast, the tip of the cylinder consists of the region with the largest deflection for all fibres. 

Both the strain at the root and the tip deflection of the cylinder reflect the overall integrated impact of the flow on the structure. Figure \ref{fig:Figure_3.1} \textit{(b)} provides an estimation of the behaviour of the time-averaged loads experienced by the cylinder, for each FST condition. $\delta_{\text{tip}}$ presents a positive correlation with $TI$, with no apparent correlation with $\mathcal{L}_{13}/D$ suggesting that $TI$ dominates the evolution of the time-averaged loads over the bluff-body. As documented in table \ref{tab:Uinf_fstcases}, the introduction of the turbulence generating grids accounted for an increase of the free-stream flow velocity introduced by an additional blockage in the water flume. Nonetheless, no correlation between the increased free-stream velocity and instrumented loads was identified.  The different dynamics introduced by the presence of FST dominated the characteristic loading regimes acting on the cylinder.

To explore the impact of the introduction of FST on the developed fluctuating loads, we start by applying a Reynolds decomposition to the captured strain signal, extracting the mean ($\overline{\varepsilon}$) and fluctuating ($\varepsilon'$) strain field:
\begin{equation}
	\varepsilon(y/D, t, \theta_f^{\alpha}) = \overline{\varepsilon(y/D, \theta_f^{\alpha})} + \varepsilon^{\prime} (y/D, t, \theta_f^{\alpha}),
\end{equation}
the latter associated with the superimposition of the effects of the flow structures shed onto the wake, and those present in the free-stream. 

An ensemble average of $rms(\varepsilon^{\prime})$ over the root region of the cylinder ($y/L_{i}\approx 0$) for the set of $4$ fibres is conducted, characterising the impact of the induced fluctuating root bending moment evolution, for the different FST conditions tested. This quantity herein referred to as $\gamma$, provides insight into the magnitude of cyclic variations of strain experienced by the cylinder, induced by the different fluctuating root bending moment dynamics it undergoes introduced by the different FST conditions.

\begin{figure}
	\centering
	\begin{subfigure}
		\centering
		\hspace*{-0.3cm}{
			\raisebox{2in}{(\textit{a})}\includegraphics[width=1\columnwidth]{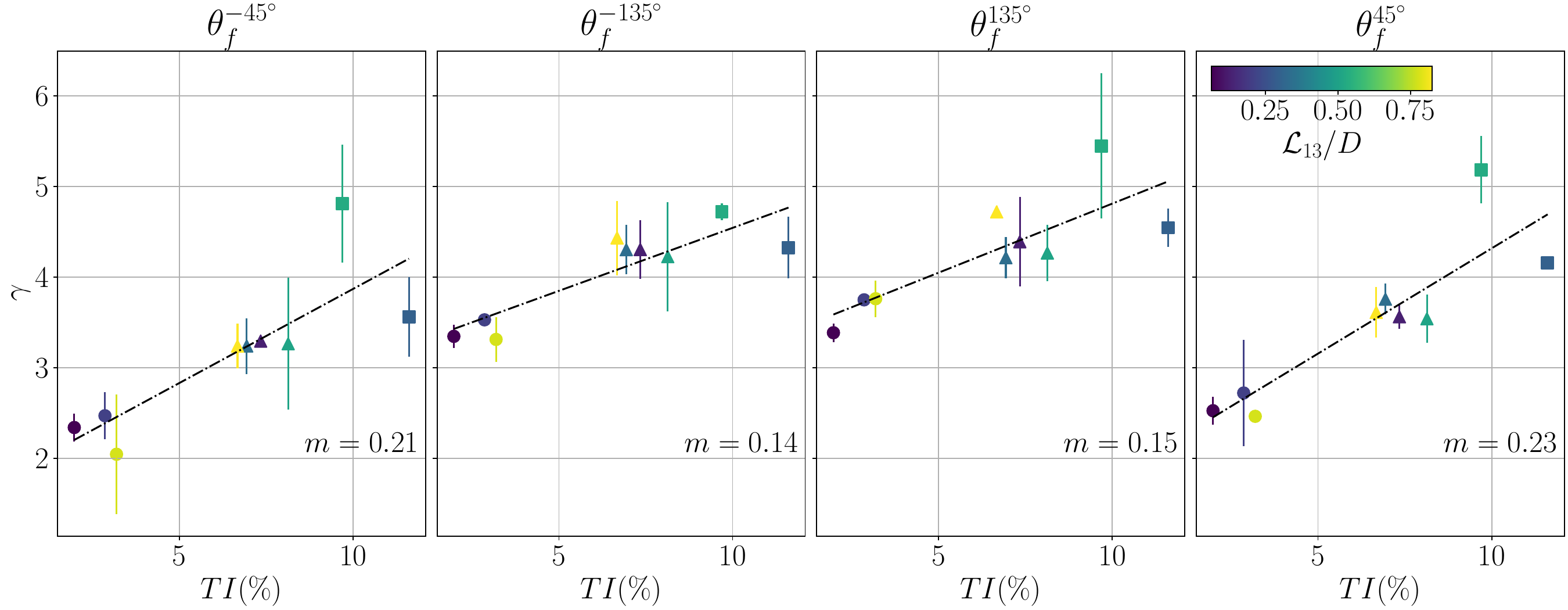}}
	\end{subfigure}
	\begin{subfigure}
		\centering
		\hspace*{-0.3cm}{
			\raisebox{2in}{(\textit{b})}\includegraphics[width=1\columnwidth]{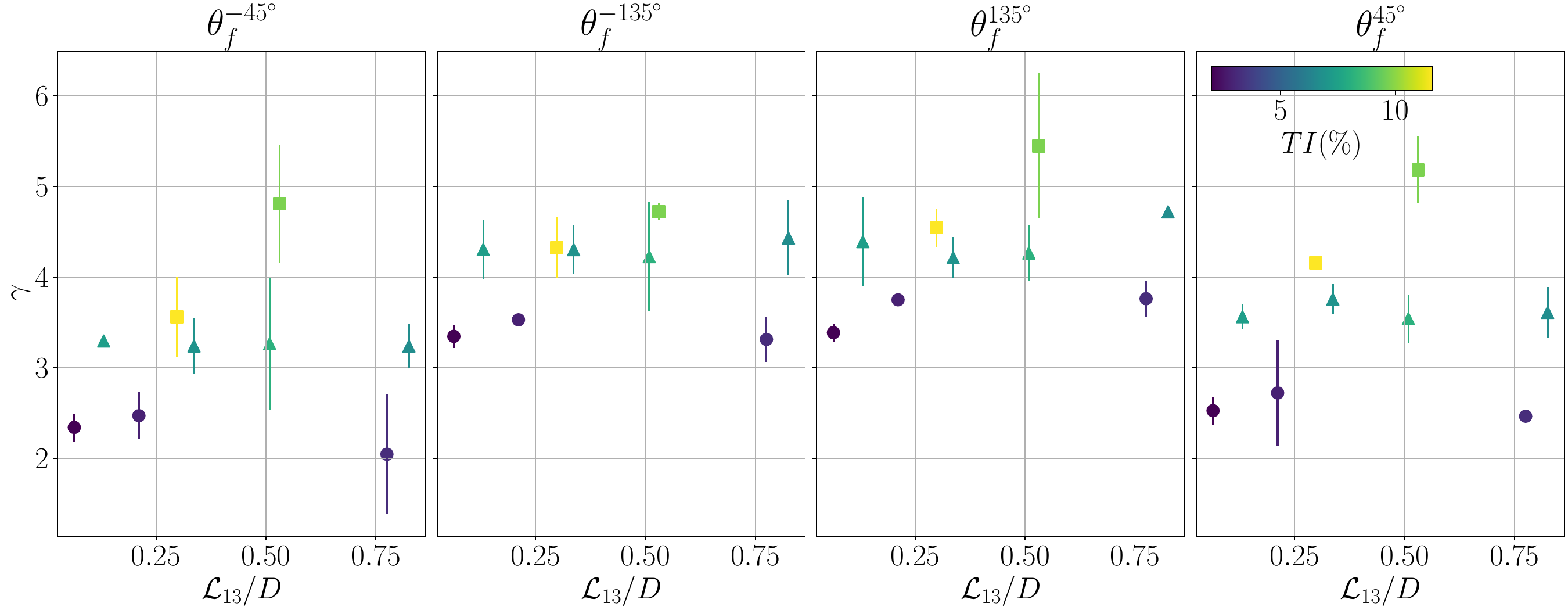}}
	\end{subfigure}
	\caption{Fluctuating root bending stresses, characterised by $\gamma$, with respect to $TI$ \textit{a)} and $\mathcal{L}_{13}/D$ \textit{b)} content in the free-stream. $m$ corresponds to the slope of the linear best fit of the evolution of $\gamma$ with $TI$.}
	\label{fig:Figure4.1}
\end{figure}

Figure \ref{fig:Figure4.1} depicts the evolution of $\gamma$, for the different $\theta_{f}^{\alpha}$, with respect to $TI$ (figure \ref{fig:Figure4.1} \textit{(a)}), and $\mathcal{L}_{13}/D$ (figure \ref{fig:Figure4.1} \textit{(b)}). The scattered points of $\gamma$ consist of the average between the two tests conducted for each FST condition, and the associated error bars correspond to the standard deviation of $\gamma$ between the two measurements taken for the same FST conditions. Both $\mathcal{L}_{13}/D$ and $TI$ present a positive correlation with the root bending $\gamma$. However, a stronger positive correlation with $TI$ is observed. The increase of $TI$ contributes to the increase of the experienced fluctuating root bending moment with an apparent linear behaviour. The dark line plotted against the scatter plot in figure \ref{fig:Figure4.1} \textit{(a)} corresponds to the best linear fit of the data spread, presenting an average slope of $m \approx 0.14$ for $\theta_f^{\pm 135^{\circ}}$, and  $m \approx 0.22$ for $\theta_f^{\pm 45^{\circ}}$. This suggests that the two faces of the cylinder are affected by different conditions, indicative of the direct and indirect impacts of FST. The increasingly energetic fluctuating velocity field impacting on the windward surface of the cylinder has a clear impact on the dynamics in this region of the cylinder, decreasing the difference in magnitude of $\gamma$ between the two surfaces for each FST case, homogenizing the structural dynamics experienced by the cylinder.

Moreover, the magnitude of $\gamma$ acquired by $\theta_{f}^{\pm 135^{\circ}}$ and $\theta_{f}^{\pm 45^{\circ}}$ is shown to be larger on the leeward side of the cylinder ($\theta_{f}^{\pm 135^{\circ}}$). This is consistent for each of the instrumented regions of the cylinder, under each FST conditions. The surface closer to the location of the shed vortices seemingly suffers from their impact. This is assumed to be a direct effect of the naturally turbulent wake of the cylinder, where the generated flow structures impact the nearby structure surface, acting locally on the surface of the cylinder. Despite $TI$ seemingly dominating the dynamics of the fluctuating root bending moment, we can also distinguish a strong secondary effect introduced by $\mathcal{L}_{13}/D$. The evolution of $\mathcal{L}_{13}/D$ in figure \ref{fig:Figure4.1} presents a positive non-monotonic correlation, with a maximum at $\mathcal{L}_{13}/D \approx 0.5$, notably half of the diameter of the cylinder. The increase of $\mathcal{L}_{13}/D$ on FST has been linked to increased energy of large-scale structures shed by the cylinders, especially in their near-wakes by promoting near-wake entrainment into the cylinder's near-wake \citep{krishna2023}. The secondary effect of $\mathcal{L}_{13}/D$ seen in this work may thus be linked to increased near-wake entrainment.

\section{Impact of the flow dynamics on the cylinder's response}\label{sec:results2}

\begin{figure}
    \centering
	\hspace*{-0cm}{\includegraphics[width=\textwidth]{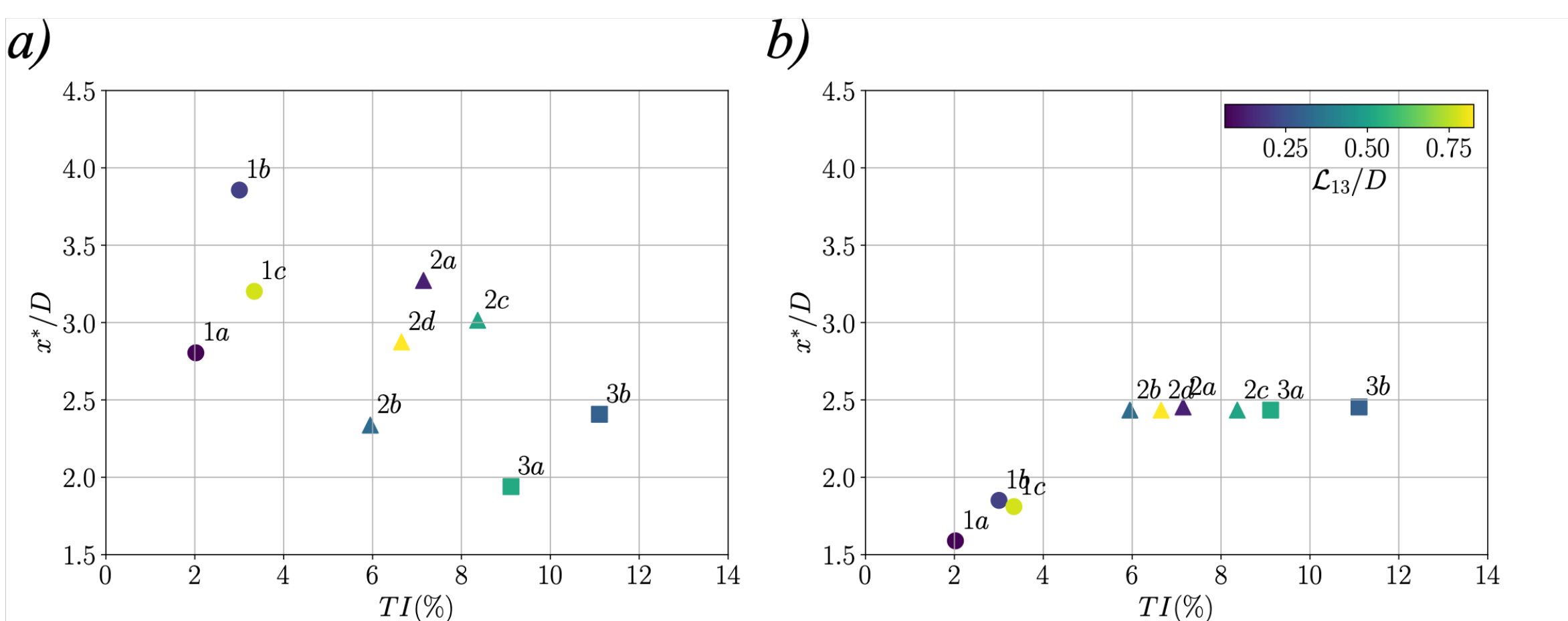}\hspace*{-0.cm}}
	\caption{\textit{a)} and \textit{b)}: vortex formation length ($x^{*}/D \rightarrow max(rms(\mathbf{u}_3^{\prime}(-1<z/D<1)))$) variation of the wake of the cylinder, for each of the FST conditions tested, and each FOV interrogated (FOV A $\rightarrow$ \textit{a)}, FOV B $\rightarrow$ \textit{b)}).}
	\label{fig:Figure3.2}
\end{figure}

The observed increase in the time-averaged loads, and magnitude of the induced fluctuating root strain introduced by FST may arise from the combination of $3$ factors:
\begin{itemize}
\item decreased vortex formation length, bringing the low-pressure vortices closer to the cylinder;
\item increased energy associated with the various flow structures;
\item increased spanwise coherence of such flow structures;
\end{itemize}
which collectively contribute to the increased magnitude of the cylinder's response. 

We will start by exploring the first point represented in figures \ref{fig:Figure3.2} - \textit{a)},\textit{b)}, by assessing the correlation between the vortex formation length in the two FOVs instrumented, and the FST conditions. The vortex formation length is defined as the streamwise location where $rms(u_3)$ reaches a maximum value, denoted by $x^{*}/D$. At the midspan of the cylinder (FOV A - figure \ref{fig:Figure3.2} \textit{a)}), $x^{*}/D$ decreases with the increase of $TI$ in the free-stream. This is consisted with \cite{gerrard1966}, and in line with \cite{krishna2023}, where the authors postulate that the increase of FST $TI$ increases entrainment via large-scale engulfment in the near-wake of the cylinder.

Figure \ref{fig:Figure3.2} \textit{b)} presents smaller values of $x^{*}/D$ for $TI<4\%$ than \textit{a)}. The reduced vortex formation length results from the introduced momentum and decreased pressure gradient added by the free-end condition at the tip of the cylinder \citep{hain2008,porteous2014,crane2021}, pulling the generated vortices closer to the cylinder. As $TI$ is increased, so is the vortex formation length for $TI < 4 \%$. However, for $TI> 6 \%$, $x^*/D$ seems to saturate at $x^*/D \approx 2.5$. The saturation length of vortex formation at FOV B is close to the vortex formation extent at the midspan of the cylinder for the same $TI$ conditions, suggesting an increased spanwise coherence of the vortices shed by the cylinder, and a diminished $3$D effect of the free end. Effectively, the presence of a similar vortex formation length at the free end of the cylinder to what is experienced at midspan suggests that both regions are subjected to similar flow physics. Similarly, $TI$ has been previously associated with an increased spanwise coherence of large-scale structures \citep{francisco2024,maryami2019}. 

Similarly to $\delta_{\text{tip}}$, $TI$ governs the evolution of the vortex formation length of the cylinder, presenting a negative correlation with $x^*/D$ for FOV A. Similarly, a negative correlation between the decrease of vortex formation length, and the increase of the time-averaged loads over the cylinder may be drawn. In the current experimental campaign, no clear correlation has been observed between $\mathcal{L}_{13}/D$ and $x^{*}/D$. The energy spectra of the fluctuating velocity fields obtained allow us to analyse the relative modification of energy associated with the flow structures shed from the cylinder into the wake, at the two spanwise locations, and assess the effect of the presence of FST. The energy spectra of the strain fluctuations along the spanwise extent of the cylinder allow us to better understand how the structural response is modified along the cylinder's extent. By analysing the structural response alongside the fluctuating velocity field's spectrum, we can attribute the structural dynamic response to specific flow structures both present in the free-stream, and the wake of the cylinder. Effectively, correlations between the energy increase of the fluctuating strain within a specific frequency band, that may be associated with a flow structure, may then be assessed from the relative energy, at the same frequency band, in the fluctuating velocity spectrum.

\begin{figure}
	\centering
	\vspace{-1cm}
	\centering\includegraphics[width=.85\columnwidth]{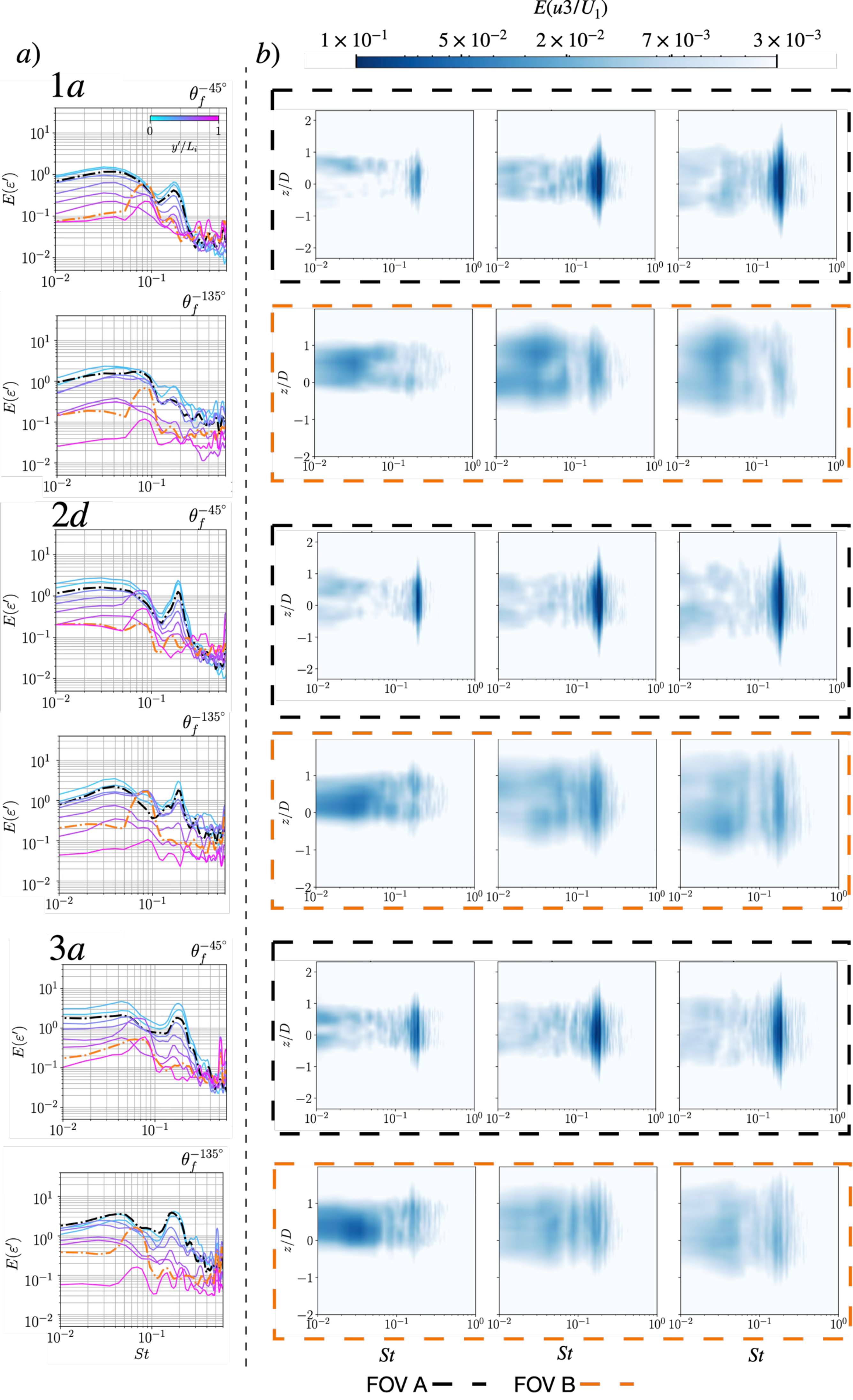}
	\caption{For FST cases \([1a, 2d, 3a]\): \textit{a)} energy spectra of the instrumented fluctuating strain \(\varepsilon'\) for \(\theta_f^{\alpha}\) with \(\alpha = [-45^{\circ}, -135^{\circ}]\) at various spanwise positions (notice how the FOV locations are highlighted); \textit{b)} energy spectra of \(u_3\) along the transverse direction of the flow at \(x/D = [0.5, 1.5, 2.5]\).}
	\label{fig:Figure6.1}
\end{figure}

The flow over a cantilevered cylinder is marked by the presence of tip vortices generated at the free end of the cylinder characterised by low frequency dynamics ($St = [6,10]\times 10^{-2}$), and a growing regular von K\'{a}rm\'{a}n vortex street towards the root \citep{porteous2014,hain2008}. Figure \ref{fig:Figure6.1} presents the energy spectra of the concurrent measurements of strain, and velocity fluctuations $u_3$ respectively obtained by the RBS (\textit{a)}) and $2$D PIV (\textit{b)}), at FOV A and B, for FST cases $[1a, 2d, 3a]$.

We start by addressing the spectra obtained from the structural response, captured by $2$ of the $4$ sets of fibre sensing lines - $\theta_f^{\alpha}$, $\alpha=[-45^{\circ}, -135^{\circ}]$. Figure \ref{fig:Figure6.1} reveals a global increase in the overall \enquote{energy} associated with the strain fluctuations of the cylinder, as the root of the body is approached ($y'/L_{i} = 0$), similarly to what has been previously evidenced in figure \ref{fig:Figure_3.1} \textit{(a)}. Furthermore, an \enquote{energy} peak at $St \approx 0.2$, associated with the von K\'{a}rm\'{a}n vortex shedding is present in every \enquote{energy} spectrum, with a relative increase of magnitude with the introduction of FST, both for the fibres on the windward and leeward region.

As we progress towards the tip of the cylinder ($y'/L_{i}=1$), we see an increased broadband energy at $St \approx 0.08$, representative of the impact of the tip vortices, shed from the free end of the cylinder \citep{porteous2014}. Moreover, the \enquote{energy} associated with the structural response introduced by these flow structures increases, with the introduction of FST.  In addition, similarly to the increased root bending moment stress ($\gamma$), the \enquote{energy} of the strain dynamics is consistently larger in the leeward face of the cylinder.

The energy spectra of the transverse velocity-component along the transverse direction of the flow, acquired in FOV A and B for FST cases $1a$, $2d$, and $3a$, show an increase of vortex shedding intensity at $St \approx 0.2$ with the introduction of FST, especially close to the cylinder. It is clear that vortex shedding is the flow structure with the largest energy in FOV A. Similarly, the energy spectra obtained from the velocity field acquired by FOV B show an increase in energy in the characteristic vortex shedding frequency, with the introduction of FST, when compared to case $1a$. This trend is seemingly not systematic with $TI$, case $2d$ presents increased energy within the vortex shedding frequency band when compared to $3a$, exposed to larger $TI$. To evaluate the impact of $\{TI, \mathcal{L}_{13}/D\}$ on the available energy for this flow structure, we analyse the streamwise evolution of relative energy of the fluctuating velocity field at the vortex shedding band $\Phi^{VS}$, defined as:

\begin{equation}
	\Phi^{VS} = \langle\int_{0.15}^{0.22} E(u_3/U_1)\mathrm{d}St\rangle_{z/D}\vert_{-1.5}^{1.5},
\end{equation}
where $\langle \Phi \rangle_{z/D}\vert_{-1.5}^{1.5}$ corresponds to the ensemble average of $\Phi$ within the range $-1.5<z/D<1.5$.

\begin{figure}
	\centering
	\vspace{-1cm}
	\centering\includegraphics[width=.95\columnwidth]{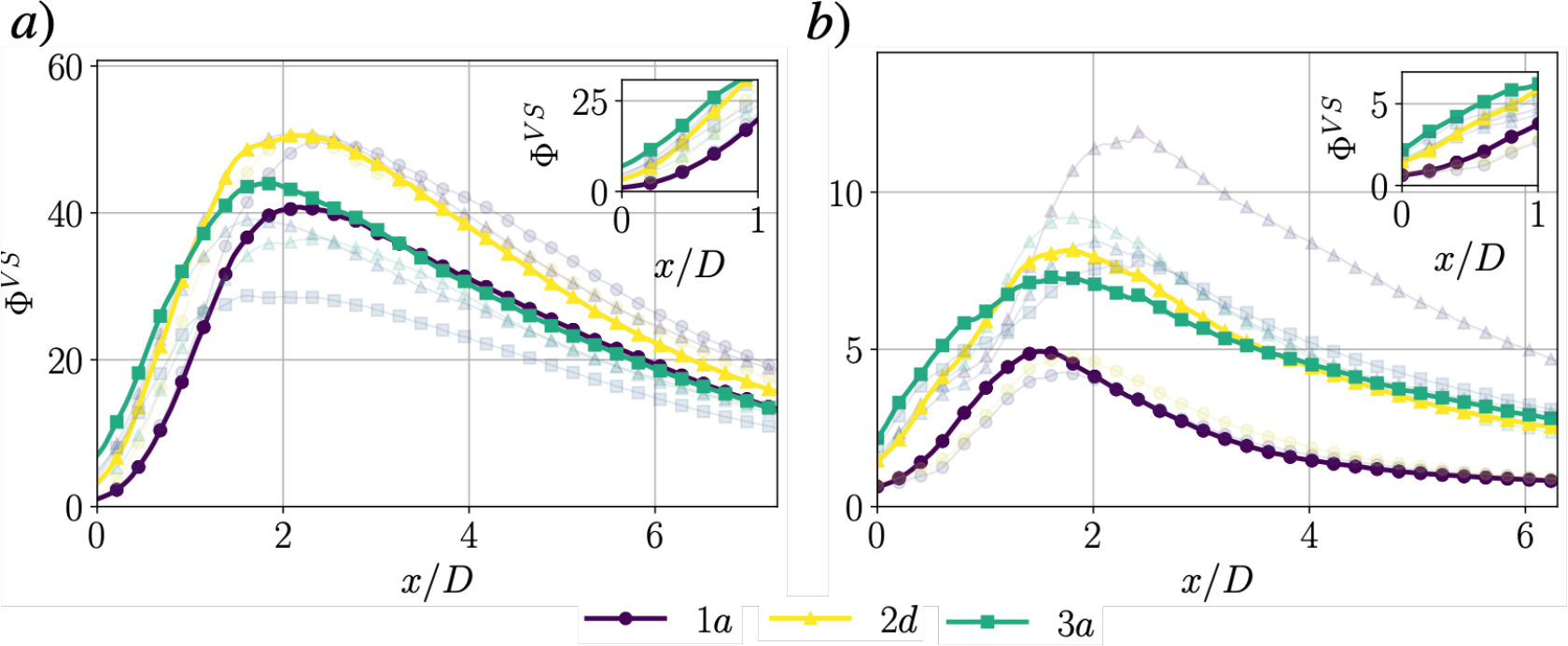}
	\caption{Streamwise evolution of the energy of $u_3/U_1$ associated with regular vortex shedding for the tested FST conditions, in FOV A (\textit{a)}) and FOV B(\textit{b)}).}
	\label{fig:energy_spectra_integration}
\end{figure}

Figure \ref{fig:energy_spectra_integration} represents the streamwise evolution of $\Phi^{VS}$ for the explored FST \enquote{flavours}, highlighting cases $[1a, 2d, 3a]$ for FOV A and B. The peak of $\Phi^{VS}$ for each FST case closely matches the vortex formation extent defined in figure \ref{fig:Figure3.2}. The increase of $TI$ results in an increase of the intensity of vortex shedding close to the cylinder (see insets of figure \ref{fig:energy_spectra_integration}). Despite presenting less energy far from the cylinder, this increased energy available at earlier streamwise stations is postulated to be transferred more efficiently to the cylinder, leading to an increased structural response induced by the corresponding flow structure.

Furthermore, tip vortices and low-frequency flow components ranging from $St = [6,10]\times 10^{-2}$ \citep{hain2008, porteous2014}, are also captured by FOV B whilst in FOV A, these structures are less prominent. This represents the cellular shedding of the cantilevered cylinder along its spanwise direction described in \cite{porteous2014}, and implies the physical origin of the increased energy at $St \approx 0.08$ of the fluctuating strain \enquote{energy} spectra closer to the tip. 

The increase in the intensity of the structural response at the characteristic Strouhal number of the regular vortex shedding reflects the modification of these coherent flow structures by the presence of FST. \cite{ramesh2024vortex} similarly reported an increase in the intensity of vortex shedding and its respective influence on the structural response of a double fixed cylinder, with the introduction of moderate levels of FST. 

Overall, the introduction of FST accounts for an increase of the relative energy associated with vortex shedding. Close to the cylinder there is an increase of energy in the regular vortex shedding frequency band, in line with the decrease of the vortex formation length seen previously in figure \ref{fig:Figure3.2} \textit{a)}. In addition, the introduction of FST suggests an increased spanwise coherence of such flow structures, as seen by the energy content presence in FOV $B$, at the regular vortex shedding characteristic $St \approx 0.2$. This corroborates the correlation between the increase in the structural response induced at the regular vortex shedding characteristic frequency, driven by the combination of a larger energy associated with the corresponding flow structure, and a stronger spanwise presence, increasing its imprint on the body.

\begin{figure}
	\centering
	\raisebox{2.3in}{\textit{a})}\includegraphics[width=\columnwidth]{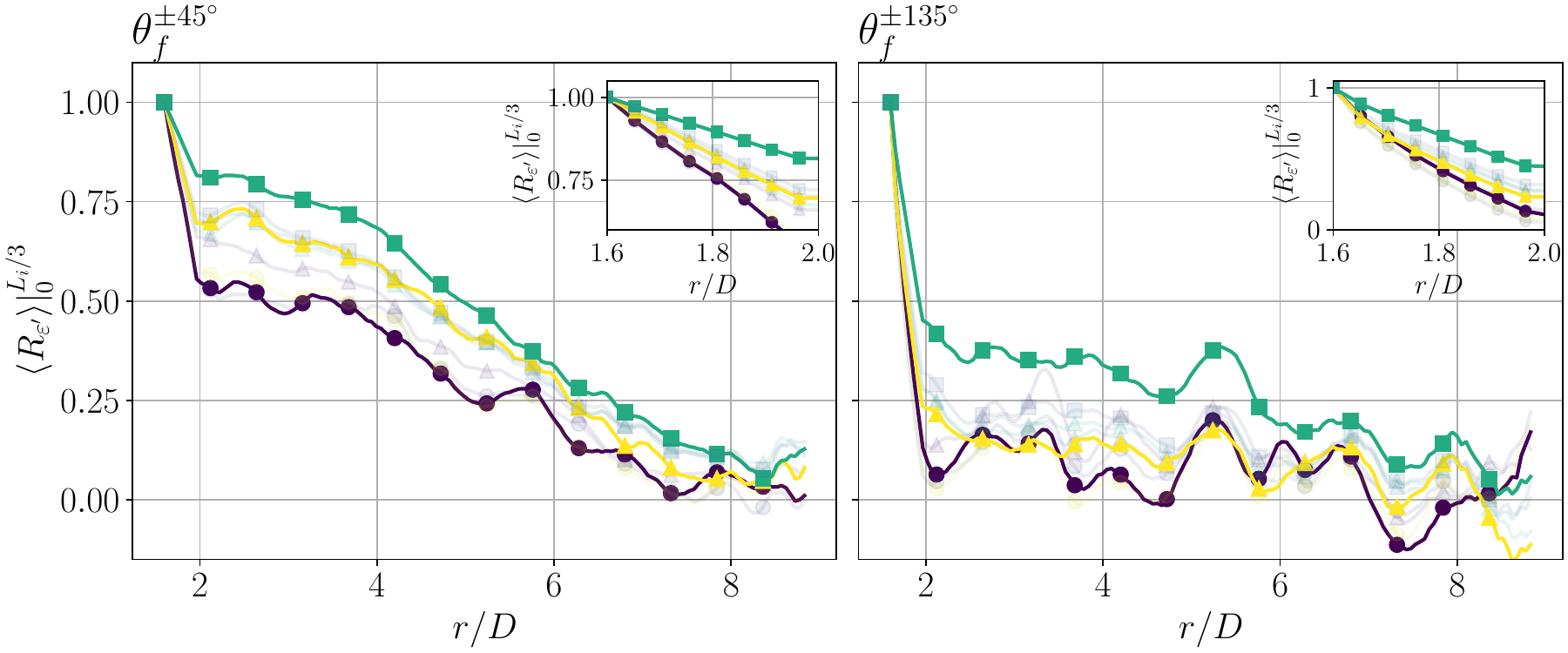}
	\raisebox{2.3in}{\textit{b})}\includegraphics[width=\columnwidth]{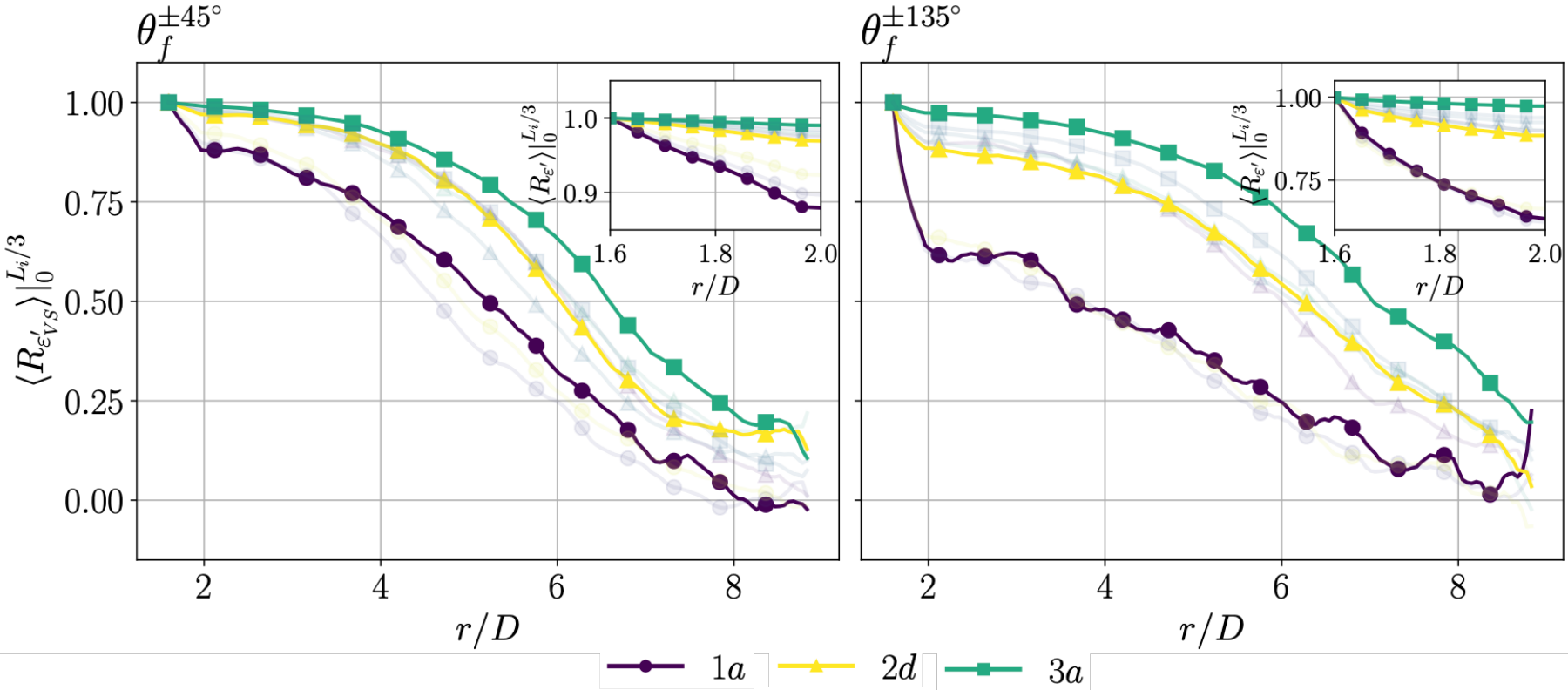}
	\caption{Autocorrelation functions $R_{\varepsilon ^{\prime}}(r/D)$ - \textit{a)} and $R_{\varepsilon ^{\prime}_{VS}}(r/D)$ - \textit{b)} along the spanwise extent of the cylinder, highlighting cases [$1a$, $2d$ and $3a$]. }
	\label{fig:Figure3.3}
\end{figure}

Having postulated an increased spanwise coherence of regular vortex shedding with the introduction of FST ultimately resulting in a larger magnitude of the strain field, we now analyse the impact of FST on the coherence of the fluctuating strain field by analysing the auto-correlation function ($R_{\varepsilon^{\prime}}(r/D)$) between ($\varepsilon^{\prime}(t, y'/L_{i})$) and ($\varepsilon^{\prime}(t, y'/L_{i} + r e_y)$), on both the windward and leeward face of the cylinder. This metric is averaged over the first $1/3$rd measured span of the cylinder starting from the root ($\langle R_{\varepsilon^{\prime}}\rangle_{y'/L_{i} \in [0, 1/3]}$), to capture the spanwise behaviour of the strain dynamics over a region close to the root of the cylinder. This metric is presented in figure \ref{fig:Figure3.3} \textit{a)}, and highlighted for cases $1a$, $2d$ and $3a$, for the windward ($\theta_{f}^{\pm 45^{\circ}}$) and leeward ($\theta_{f}^{\pm 135^{\circ}}$) faces of the cylinder. Figure \ref{fig:Figure3.3} \textit{b)} shows $\varepsilon^{'}_{VS}$, computed from the raw  $\varepsilon^{'}$ but with a bandpass filter isolating the induced strain by regular vortex shedding within $St = [0.15, 0.22]$, allowing us to analyse how this flow structure impacts the structural dynamics along the spanwise extent of the cylinder.

The different evolution of $\langle R_{\varepsilon^{\prime}}\rangle_{y'/L_{i} \in [0, 1/3]}$ in the two faces as presented in figure \ref{fig:Figure3.3} \textit{a)} reflects the different dynamics imposed on the two faces of the cylinder, before and after the shear-layer separation points. This is the case even without the introduction of FST, evidencing the aforementioned relevance of the direct/indirect impact of FST. Strain fluctuations become uncorrelated over shorter distances (along the spanwise direction) on the leeward side of the cylinder and contrarily, the windward side sustains correlated strain fluctuations for larger regions of the cylinder. Furthermore, the decreased magnitude of $\langle R_{\varepsilon^{\prime}(r/D)}\rangle_{y'/L_{i} \in [0, 1/3]}$ as $r/D$ increases shows that the fluctuating strain acquired at larger distances between two points becomes increasingly uncorrelated. 

By using a bandpass filter to isolate the regular-vortex shedding influence on the fluctuating strain field, we assess the direct impact of this flow structure on the coherence of $\varepsilon^{'}$ over the two faces of the cylinder, as presented in figure \ref{fig:Figure3.3} \textit{b)}. For the same spanwise location and same FST conditions, $\langle R_{\varepsilon^{\prime}_{VS}}\rangle_{y'/L_{i} \in [0, 1/3]}$ presents lower values on the leeward face when compared to the windward face, reflecting the impact of the proximity of the flow structure to the surface under analysis. By comparing the evolution of $\langle R_{\varepsilon^{\prime}}\rangle_{y'/L_{i} \in [0, 1/3]}$ on the windward side of the cylinder presented in figures \ref{fig:Figure3.3} \textit{a)} and \textit{b)}, we can see that regular vortex shedding largely contributes to the spatial dynamics of the induced fluctuating strain, even when FST is introduced.

Overall, the introduction of FST accounts for a slower decay of autocorrelation for both the windward and leeward face strain dynamics, and increased coherence of strain fluctuations. The introduction of FST increased the spanwise coherence of strain induced by vortex shedding in the windward and leeward faces, allowing us to indirectly infer an increased coherence of the corresponding flow structure. The different behaviour of the correlation between $\theta_f^{\pm 45^{\circ}}$ and $\theta_f^{\pm 135^{\circ}}$ highlights that flow structures generated in the immediate wake of the cylinder impact the experienced loads differently on both faces.

Having analysed the impact of FST on the spatial autocorrelation of the strain fluctuations, we now characterise its impact on the temporal strain dynamics on the leeward and windward sides. We focus on $\tau_{\varepsilon^{\prime}}$, defining the characteristic time over which strain fluctuations remain correlated. $\tau_{\varepsilon^{\prime}}$ is then defined as:

\begin{equation}
	\tau_{\varepsilon^{\prime}}(y'/{L_{i}}, \theta_f^{\alpha}) = \big{\langle}\int_0^{\tau_0}R_{\varepsilon^{\prime}}(\tau)\mathrm{d}\tau \big{\rangle}_{y'}\big{\vert}_{0}^{L_{i}/3},
	\label{tau_epsilon}
\end{equation}
where $R_{\varepsilon^{\prime}}(\tau)$ corresponds to the autocorrelation function between $\varepsilon^{\prime}(t, y'/L_{i}, \theta_f^{\alpha})$ and $\varepsilon^{\prime}(t + \tau, y'/L_{i}, \theta_f^{\alpha})$, $\tau_0$ corresponds to the first zero crossing position over the time lag ($\tau$) of the autocorrelation function, as represented in figure \ref{fig:Figure3.4} \textit{a)} for FST cases $1a$, $2d$ and $3a$ and $\langle \cdot \rangle_{y'}\vert_{0}^{L_i/3}$ to the ensemble average of $\tau_0$ over the first measured third of the cylinder, i.e. $0 \leq y^{\prime} \leq L_{i}/3$ 

\hspace{1cm}{
\begin{figure}
	\centering{	
	\hspace*{-2.0cm}{\raisebox{0.1in}{\raisebox{2.1in}{(\textit{a})}\includegraphics[width=.5\columnwidth]{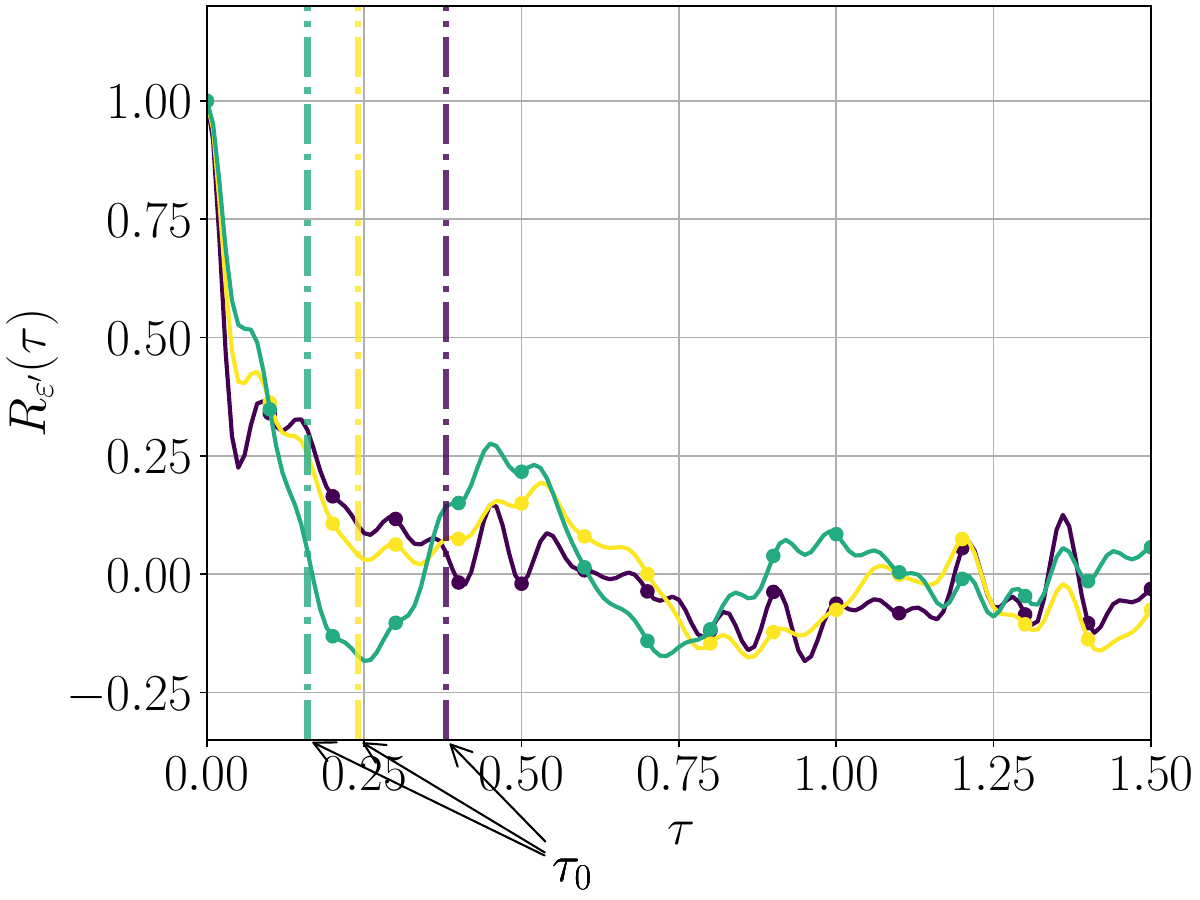}}}
	\raisebox{2.1in}{(\textit{b})}\includegraphics[width=.55\columnwidth]{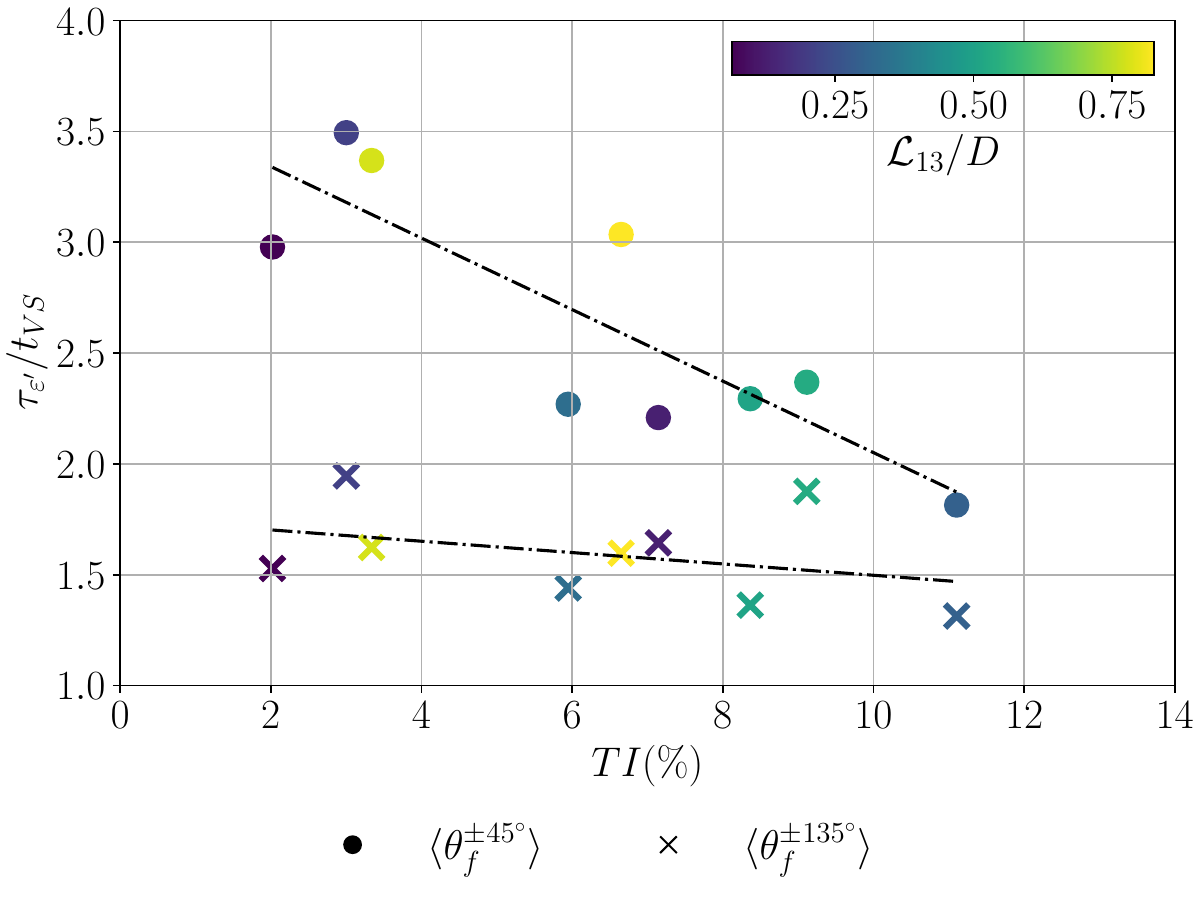}}
	\caption{\textit{a}): Autocorrelation function of $R_{\varepsilon ^{\prime}}(\tau)$ along the spanwise extent of the cylinder, for $1a$ (\raisebox{0.5ex}{\textcolor[HTML]{654C72}{\rule{0.25cm}{1pt}}}), $2d$ (\raisebox{0.5ex}{\textcolor[HTML]{EBE57E}{\rule{0.25cm}{1pt}}}) and  $3a$ (\raisebox{0.5ex}{\textcolor[HTML]{8FA798}{\rule{0.25cm}{1pt}}}). \textit{b}): autocorrelation temporal scale of the fluctuating strain signal $\tau_{\varepsilon^{\prime}}/t_{VS}$ normalised by the characteristic vortex shedding temporal scale, as a function of $TI$ experienced by the cylinder under each FST case considered, averaged between the two sensing lines on the windward and leeward face of the cylinder.}
	\label{fig:Figure3.4}
\end{figure}}

Figure \ref{fig:Figure3.4} \textit{b)} presents $\tau_{\varepsilon^{\prime}}/t_{VS}$ as a function of  $TI$ in each FST case, captured by the windward and leeward faces of the cylinder. The increase of $TI$ in the free-stream accounts for a decrease of $\tau_{\varepsilon^{\prime}}/t_{VS}$ both in the leeward and windward faces. This decrease of $\tau_{\varepsilon^{\prime}}/t_{VS}$ suggests that as $TI$ increases, the structure experiences changes in fluctuating forces faster, indicating a shorter correlation time under higher turbulence intensity conditions. High-TI environments may then lead to more frequent stress cycles, thereby accelerating wear and failure, especially in the leeward face of the structure. This decrease is steeper in the windward face of the cylinder, where the direct influence of the FST is felt. In addition, the difference between $\tau_{\varepsilon^{\prime}}/t_{VS}$ captured in the windward and leeward faces decreases with the increase of $TI$ in the free-stream, suggesting that $TI$ tends to homogenize the temporal behaviour of the structural response of the cylinder on both faces, indicating a convergence in their temporal characteristics. Moreover, the increase of $\mathcal{L}_{13}/D$ doesn't seem to influence the temporal correlation time of the strain fluctuations in both faces.

The increased impact on the loading events over the cylinder is then attributed to the combination of the $3$ effects aforementioned, rising from an increased proximity of the vortex formation region to the cylinder, increased energy available in the near-wake region immediately adjacent to the cylinder, and increased spanwise coherence of the near-wake flow structures, resulting in a larger coherence of loads acting on the cylinder. These $3$ act simultaneously, adding up to an increased effective root bending moment experienced by the cylinder, and increased energy associated with regular vortex shedding (figure \ref{fig:Figure6.1}). 

\section{Cross-correlating the flow and structural response}\label{sec:results3}

The cross-power spectral density (CPSD) between the flow data and the strain data at the concurrent $y/D$ location is now addressed, to analyse the coherence between both fields. As the measurements were concurrent, this allows us to evaluate the spatio/temporal correlation between these two signals. The CPSD is then defined as:
\begin{equation}
	CPSD(x/D,z/D,St) = \mathcal{F}(R_{\varepsilon^{\prime} u_{3} }(x/D,z/D, \tau)),
	\label{eq:CPSD}
\end{equation}
where $\mathcal{F}$ corresponds to the Fourier transform operator, and $R_{\varepsilon^{\prime} u_{3} }(x/D,z/D, \tau)$ to the cross correlation function between $\varepsilon'$ and $u_{3}$, defined as:
\begin{equation}
	R_{\varepsilon^{\prime} u_{3}^{\prime} }(x/D,z/D, \tau) = \int_{\infty}^{\infty}\varepsilon^{\prime}(t)u_{3}^{\prime}(x/D,z/D,t + \tau) \mathrm{d}t.
	\label{eq:correlation}
\end{equation}

\begin{figure}
	\centering
	\begin{subfigure}
		\centering
		\hspace*{-0.5cm}{
		\raisebox{3.5in}{(\textit{a})}\includegraphics[width=.9\columnwidth]{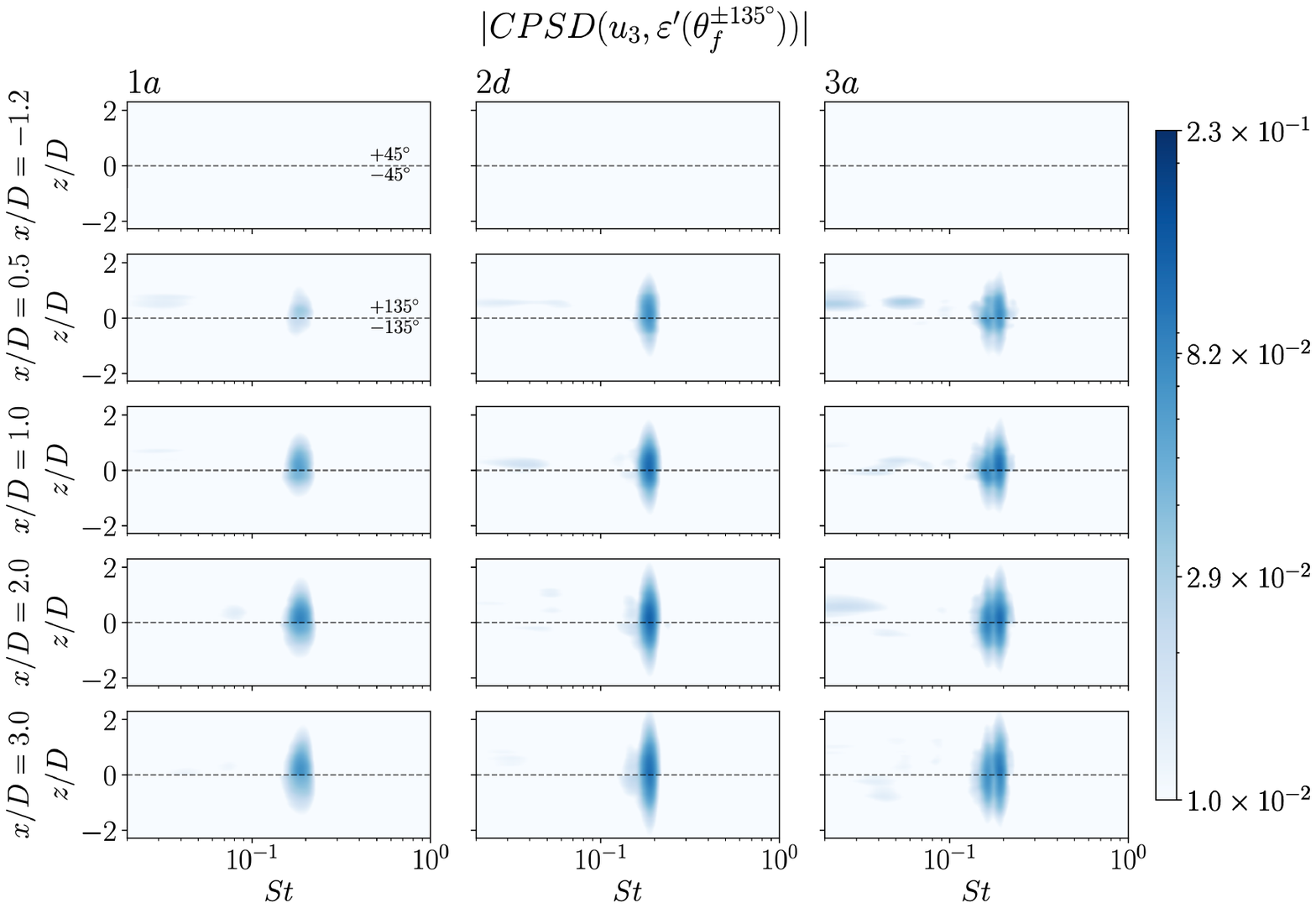}}
	\end{subfigure}
	\centering
	\begin{subfigure}
		\centering
		\hspace*{-0.5cm}{
		\raisebox{3.5in}{(\textit{b})}\includegraphics[width=.9\columnwidth]{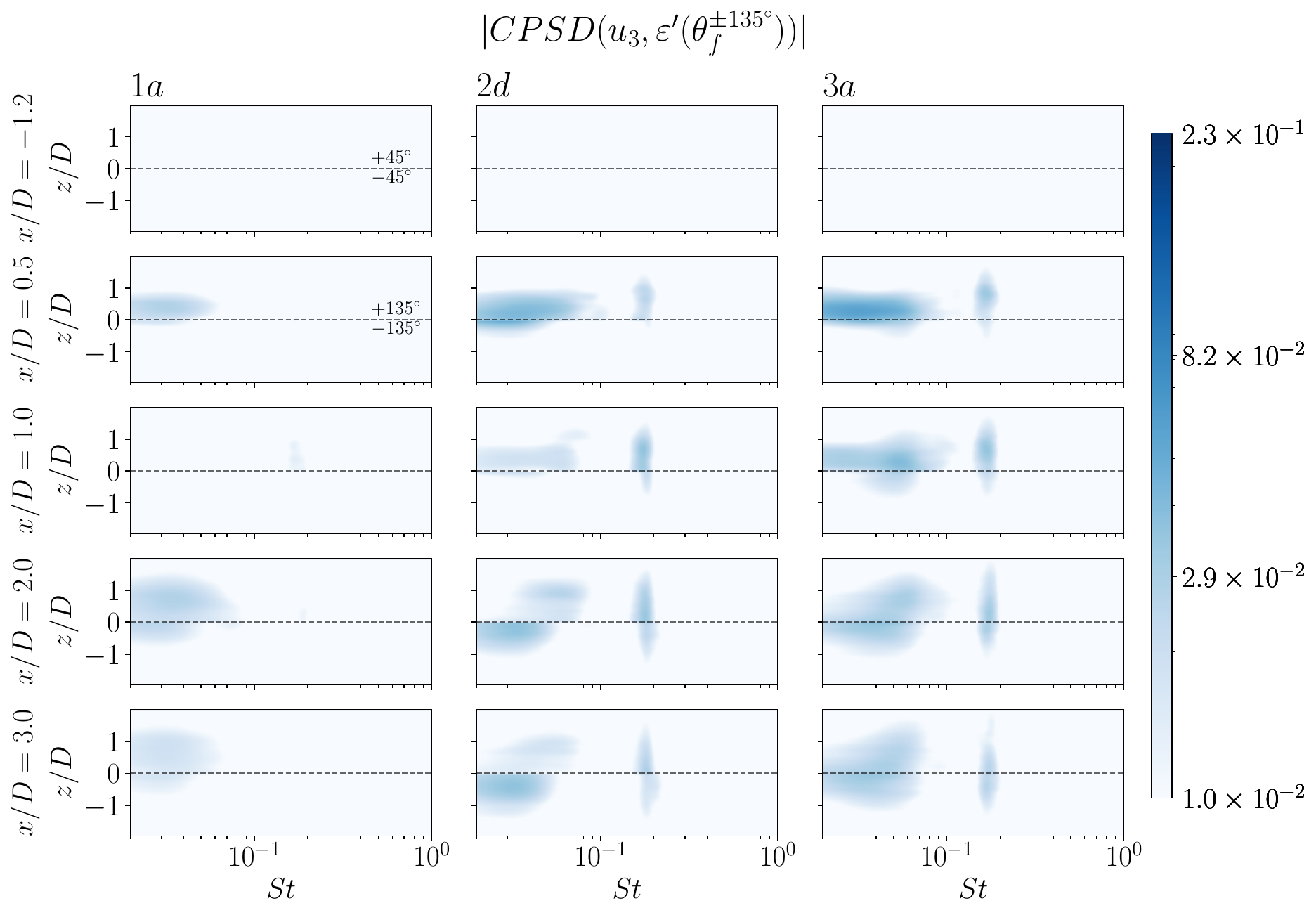}}
	\end{subfigure}
	\caption{Magnitude of the cross power spectral densities between $\varepsilon^{\prime}$, and $u_3$, for cases $1a$, $2d$, and $3a$ with velocity information extracted from FOV A (\textit{a)}) and B (\textit{b)}). The correlation is performed with the induced strain at the FOV spanwise location. For $x/D<0$, the windward fibres are used to compute the cross-correlation.}
	\label{fig:Figure6.2}
\end{figure}

Figure \ref{fig:Figure6.2} represents the magnitude of the CPSD between $u_3(z/D)$ in both FOVs A and B, at streamwise stations $x/D = [-1.2, 0.5, 1.0, 2.0, 3.0]$, and $\varepsilon^{\prime}$ at the respective $y/D$ location for each FOV, for FST cases $1a$, $2d$ and $3a$. For $z/D>0$, the CPSD is computed with the fluctuating strain field captured by $\theta_{f}^{+135^{\circ}}$, and for $z/D<0$ with $\theta_{f}^{-135^{\circ}}$. 

The magnitude of the CPSD at $x/D = -1.2$, representative of a location immediately upstream of the cylinder's position, in both FOVs is deprived of energy when compared to the CPSD contours obtained from downstream of the cylinder, even with the introduction of FST. This station shows that there is no strong coherence between the experienced loads, and incoming flow structures. Figure \ref{fig:Figure6.2} \textit{b)} shows an increase in \enquote{energy} associated with the tip vortices shed by the cylinder, reflecting both the increased energy and spanwise coherence of these structures. Furthermore, the presence of FST accounts for an increase of the energy associated with regular vortex shedding, similar to what has been discussed for the energy spectra presented in figures \ref{fig:Figure6.1}. This is true for both FOV A (figure \ref{fig:Figure6.2} \textit{a)}) and B (figure \ref{fig:Figure6.2} \textit{b)}), supporting the claim that:

\begin{itemize} 
	\item the introduction of FST accounts for an increase of spanwise coherence of large-scale structures associated with vortex shedding in a cantilevered cylinder, despite the $3$D disruption of regular shedding experienced at the tip;
	\item FST introduces more energy in the large-scale structures (as seen in the \enquote{energy} spectra of the fluctuating velocity field in figure \ref{fig:Figure6.1}), increasing the corresponding structural dynamics.
\end{itemize}

These effects combined result in an overall increased impact of the regular vortex shedding characteristic flow structures on the structural response of cylinder. In addition, the small-scales introduced by FST do not induce a coherent structural response within the cylinder. However, the structural response is indeed more susceptible to the generation, modification and deformation of large coherent structures shed from the cylinder, observable both with the acquired CPSD between the windward and leeward fibre sensors. Based on the results provided by the CPSD, the cylinder response acts as a lowpass filter, primarily responding to the flow's low-frequency content and shed structures. 

The magnitude of the CPSDs computed with the fluctuating velocity field at different streamwise stations follows a similar trend to what has been assessed before in the evolution of the flow spectra. As streamwise distance from the leeward face of the cylinder increases, after reaching the maximum, the CPSD decreases in magnitude corresponding to the energy decay of the associated flow structures impacting on the cylinder's structural response. 

The CPSDs between the strain and velocity fluctuations allowed us to examine how coherent the two fields are, and how FST affects this coherence. To assess how FST affects the temporal interaction between the convected flow structures and the cylinder's structural response, we analyse the cross-correlation between the two quantities defined in equation \ref{eq:correlation}, focusing on the documented region downstream of the cylinder ($x/D>0$) and ensemble averaging the correlation between the two fibres located on the leeward face of the cylinder ($\langle \cdot \rangle \alpha \vert _{-135^{\circ}}^{135^{\circ}}$), and over the transverse direction of the flow between $-1.5<z/D<1.5$ ($\langle \cdot \rangle_{z_D}\vert_{-1.5}^{1.5}$): $\big{\langle}\big{\langle}R_{\varepsilon^{\prime} u_{3}^{\prime} }(x/D,z/D, \tau)\big{\rangle}_{\alpha}\big{\vert}_{-135^{\circ}}^{135^{\circ}}\big{\rangle}_{z/D}\big{\vert}_{-1.5}^{1.5}$. The ensemble average over the two leeward fibres is henceforth omitted for simplicity.

 \begin{figure}
	\centering
	\includegraphics[width=\columnwidth]{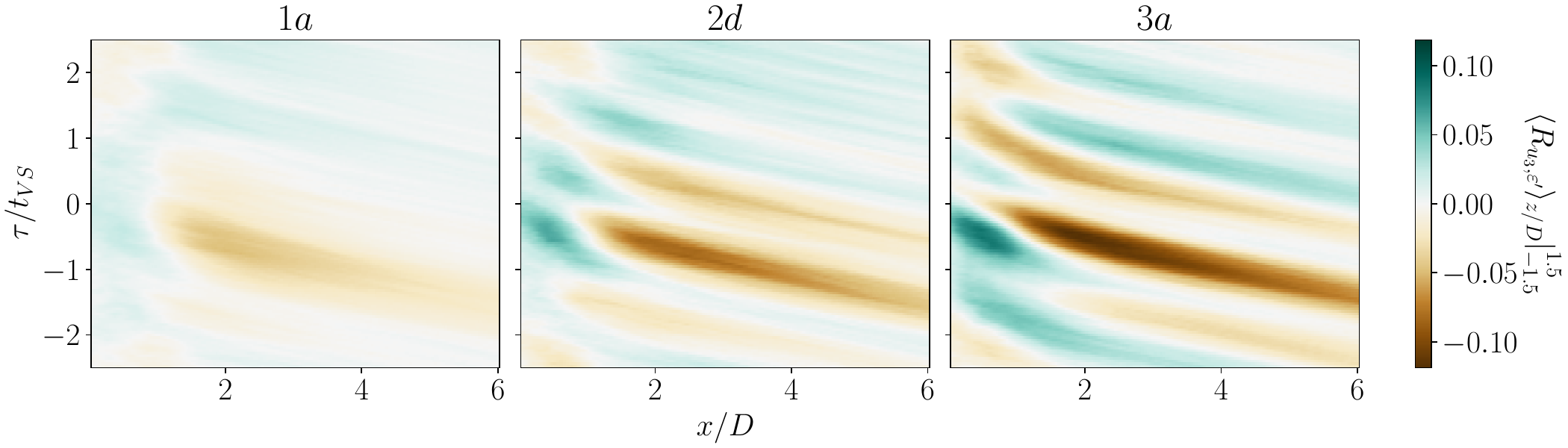}
	\caption{Evolution of $\big{\langle}R_{\varepsilon^{\prime} u_{3}^{\prime} }(x/D, \tau)\big{\rangle}_{z/D}\big{\vert}_{-1.5}^{1.5}$ along the streamwise direction, for FST cases $1a$, $2d$ and $3a$. The time lag of the correlation function $\tau$ is normalised with the vortex shedding temporal scale $t_{VS}$ relative to each FST case.}
	\label{fig:Figure6.3}
\end{figure}

Figure \ref{fig:Figure6.3} presents the streamwise evolution of $\big{\langle}R_{\varepsilon^{\prime} u_{3}^{\prime} }(x/D,z/D, \tau)\big{\rangle}_{z/D}\big{\vert}_{-1.5}^{1.5}$, for FST cases $1a$, $2d$ and $3a$. The increase of $TI$ induces a clear increase of the correlation magnitude. The periodicity in the time lag ($\tau/t_{VS}$) of the correlation persists across all cases (because vortex shedding is the dominant flow structure), but with increasing $TI$, the represented structures become sharper and more evident. This again highlights that the increase of $TI$ in FST increases the efficiency of the flow at inducing a structural response on the cylinder. 

A clear transition from regions of positive to negative correlation magnitudes at fixed streamwise positions as a function of time lag ($\tau/t_{VS}$) can also be seen in figure \ref{fig:Figure6.3}, reflecting the periodic relationship between the strain response and the flow velocity fluctuations downstream. The positive correlation peak at small $x/D$ is indicative of an initial region of the flow in which the fluctuating strain and velocity field are in phase, leading to a positive correlation between the two signals. As the $x/D$ increases, the correlation peak becomes negative, whereby the two signals become out of phase, and anti-correlated. This behaviour is not affected by the presence of FST, and consistent across all the tested cases.

Furthermore, the strain temporally leads the flow as the correlation peaks repeatedly occur at negative time-lags, for all tested cases. As $TI$ increases, this positive-to-negative transition becomes sharper and more defined and the correlation magnitude larger, reinforcing the idea that the strain response becomes more closely tied to the flow structures. This further supports the notion that turbulence amplifies the interaction between the flow and the structural response of the cylinder.

\section{Contribution of regular vortex shedding as the main coherent flow structure to induced root bending stresses }\label{sec:results4}

With the CPSDs presented in figure \ref{fig:Figure6.2}, regular vortex shedding and its modification by the presence of FST has been shown to be the most relevant flow feature contributing to the correlation between the flow and the respective structural dynamics. To assess the influence of FST on the modification of vortex shedding, and its respective influence on the retrieved structural dynamics, we isolate the flow structure in question. This is done by using a modal decomposition technique to identify the vortex shedding mode(s). We are then able to project the PIV velocity fields onto these mode(s) and thereby compute a new set of CPSDs between the fluctuating strain and the vortex-shedding flow structures. In this way we are able to isolate the contribution from vortex shedding to the overall CPSD, and hence indirectly observe the contribution to the CPSDs from the remaining flow structures present within the wake. We choose to use optimal mode decomposition using 200 ranks (OMD) \citep{Wynn2013OMD}, to extract the vortex shedding mode(s).

\begin{figure}
	\centering
	\includegraphics[width=\columnwidth]{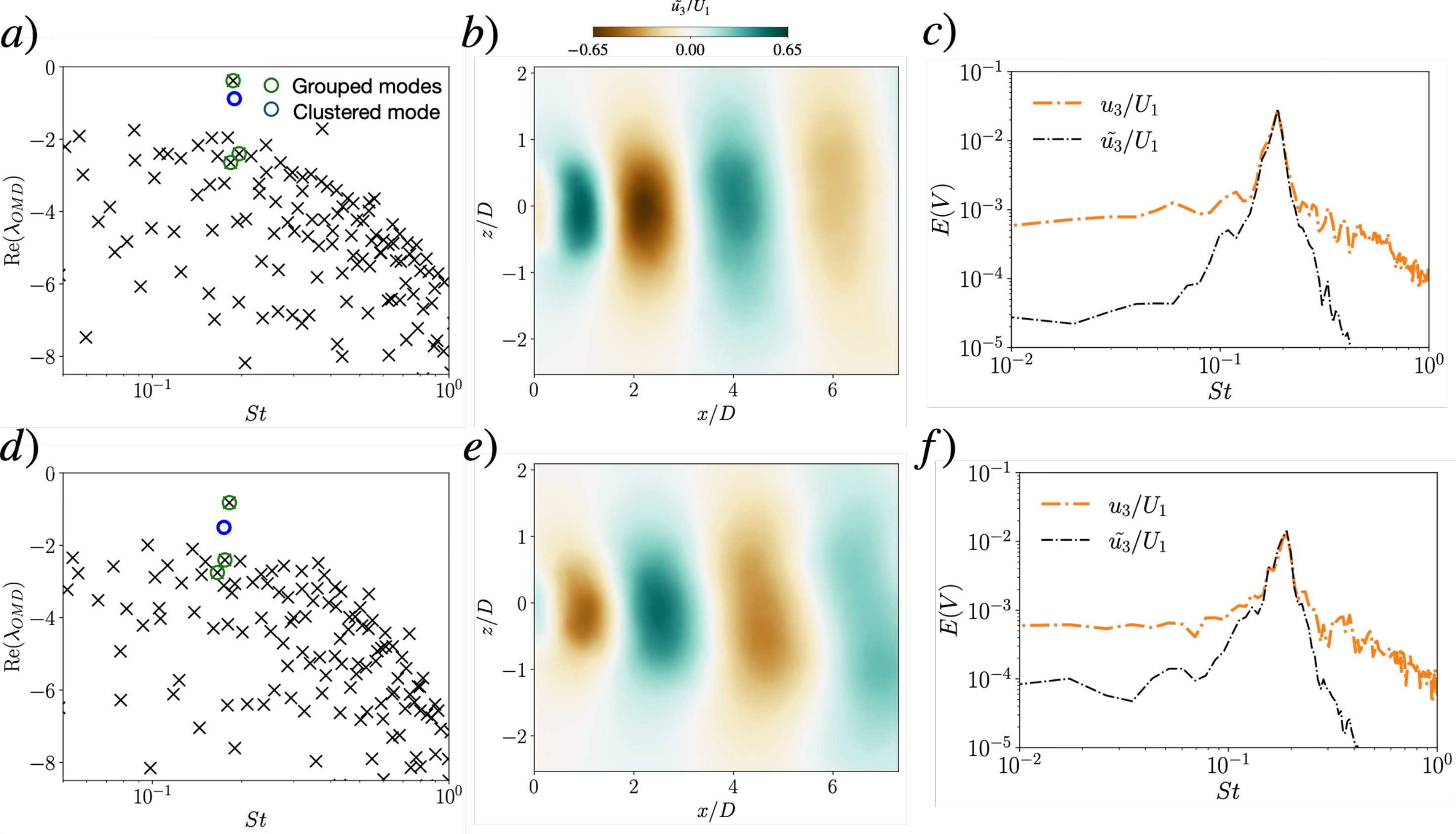}
	\caption{For FST cases $1a$ \{\textit{a)}, \textit{b)}, \textit{c)}\} and  $3a$ \{\textit{d)}, \textit{e)}, \textit{f)}\}, figures \{\textit{a)} and \textit{d)}\} represent the raw OMD modes, selected and final clustered modes after application of the clustering algorithm \citep{beit2021}; figures \{\textit{b)} and \textit{e)}\} represent the reconstructed fluctuating velocity fields $\tilde{u_3}/U_1$ at a snapshot $t$, induced by the coherent motion represented by the clustered mode, and figures \{\textit{c)} and \textit{f)}\} present the energy spectrum of reconstructed field $\tilde{u_3}/U_1$, and the raw PIV velocity field.}
	\label{fig:Figure7.1}
\end{figure}

The presence of FST broadens the frequency band associated with regular vortex shedding as seen from figure \ref{fig:Figure6.1}, and in \cite{martin2024, ramesh2024vortex, rind2012}. OMD yields a set of individual modes, with similar, but ultimately slightly different characteristic eigenvalues in the frequency range occupied by the vortex shedding. We employ a mode-clustering algorithm to group the relevant modes to reconstruct the most energetic flow structure under analysis. This algorithm is explored in \cite{beit2021} and briefly described in appendix \ref{appendix}, and exploits the spatio/temporal similarity and respective energy of the individual modes, to select and cluster them into a global coherent flow structure. Figure \ref{fig:Figure7.1} \textit{a)} and \textit{d)} present the real part and characteristic frequency of the clustered OMD mode's eigenvalues for FST cases $1a$ and $3a$ and the selected and clustered mode representing regular vortex shedding. In the same figure, \textit{b)} and \textit{e)} represent the reconstructed velocity field produced by the captured coherent flow motion, and \textit{c)} and \textit{f)} the quality of the mode clustering reconstruction, by comparing the reconstructed field $\tilde{u}_3/U_1$ and the raw PIV velocity field $u_3^\prime/U_1$. We focus this analysis on the location of FOV A at the midspan, where regular vortex shedding remains relatively unaffected by $3$D events introduced by the free end of the cylinder.

We now define $\Gamma$:
\begin{equation}
	\Gamma (x/D, y/D) = \int_{-\infty}^{\infty} \int_{-2}^{2} CPSD(V(x/D,z/D,t), \varepsilon^{\prime}(y/D,t)) \mathrm{d}(z/D) \mathrm{d}St,
\end{equation}
where $V$ consists of either the raw velocity field or the velocity field yielded through projection onto the clustered OMD vortex shedding mode.

\begin{figure}
	\centering
	\includegraphics[width=\columnwidth]{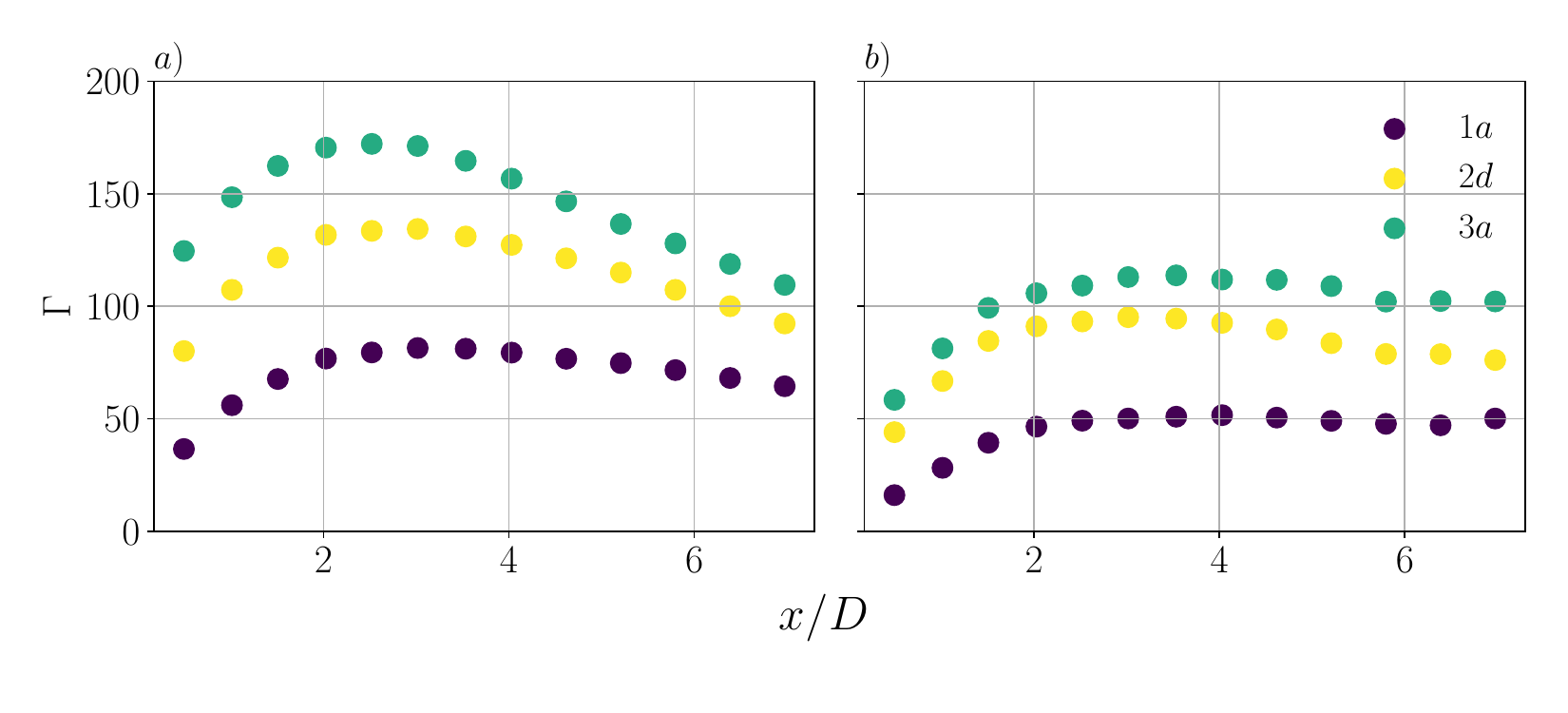}
	\caption{Influence of FST on the evolution of $\Gamma$ computed with the raw PIV velocity field $u_3/U_1$ (\textit{a)}), and the filtered velocity field by the OMD clustering $\tilde{u_3}/U_1$ (\textit{b)}), and the fluctuating strain at the FOV A location, similarly to the analysis in figure \ref{fig:Figure6.2}.}
	\label{fig:Figure7.2}
\end{figure}

Figure \ref{fig:Figure7.2} demonstrates the impact of free-stream turbulence (FST) on the evolution of $\Gamma$, calculated from both the raw PIV velocity field (\textit{a)}) which contains all the flow dynamics, including coherent and incoherent flow structures, and the OMD-filtered field (\textit{b)}) isolating vortex shedding. In figures \ref{fig:Figure7.2} \textit{a)} and \textit{b)}, $\Gamma$ increases with downstream distance $x/D$ attaining a maximum at the vortex formation extent, indicating that the influence of coherent flow structures, such as vortex shedding, becomes more prominent as they develop in the wake of the cylinder.

The increase of $TI$ leads to a larger magnitude of $\Gamma$ computed from the raw, and the filtered velocity field. From the CPSDs analysed in figure \ref{fig:Figure6.2}, we have shown that regular vortex shedding is the key flow motion interacting with the structure. However, by filtering this motion and analysing the evolution of $\Gamma$ along the streamwise direction of the flow, we can indirectly see that especially in the near-wake, incoherent motions introduced into the wake largely contribute to the \enquote{energy} associated with the velocity-strain correlation. As we progress in the streamwise direction, the magnitude of $\Gamma$ in figure \ref{fig:Figure7.2} \textit{a)} approximates the magnitude of gamma in \ref{fig:Figure7.2} \textit{b)}, representing the decay of the incoherent motions, allowing regular vortex shedding to become the dominant flow structure contributing to the structural dynamics. 

To determine the contribution of regular vortex shedding on the previously assessed increase in bending stresses, we analyse the previously defined metric $\gamma$, applying a bandpass filter within $St = [0.15, 0.22]$ to the signal.  This allows us to isolate the contribution of regular vortex shedding $\gamma^{VS}$ to $\gamma$, and compare it to the present incoherent flow motions. 

\begin{figure}
	\centering
	\hspace{-2cm}{\raisebox{2in}{\textit{a})}
	\includegraphics[width=.55\columnwidth]{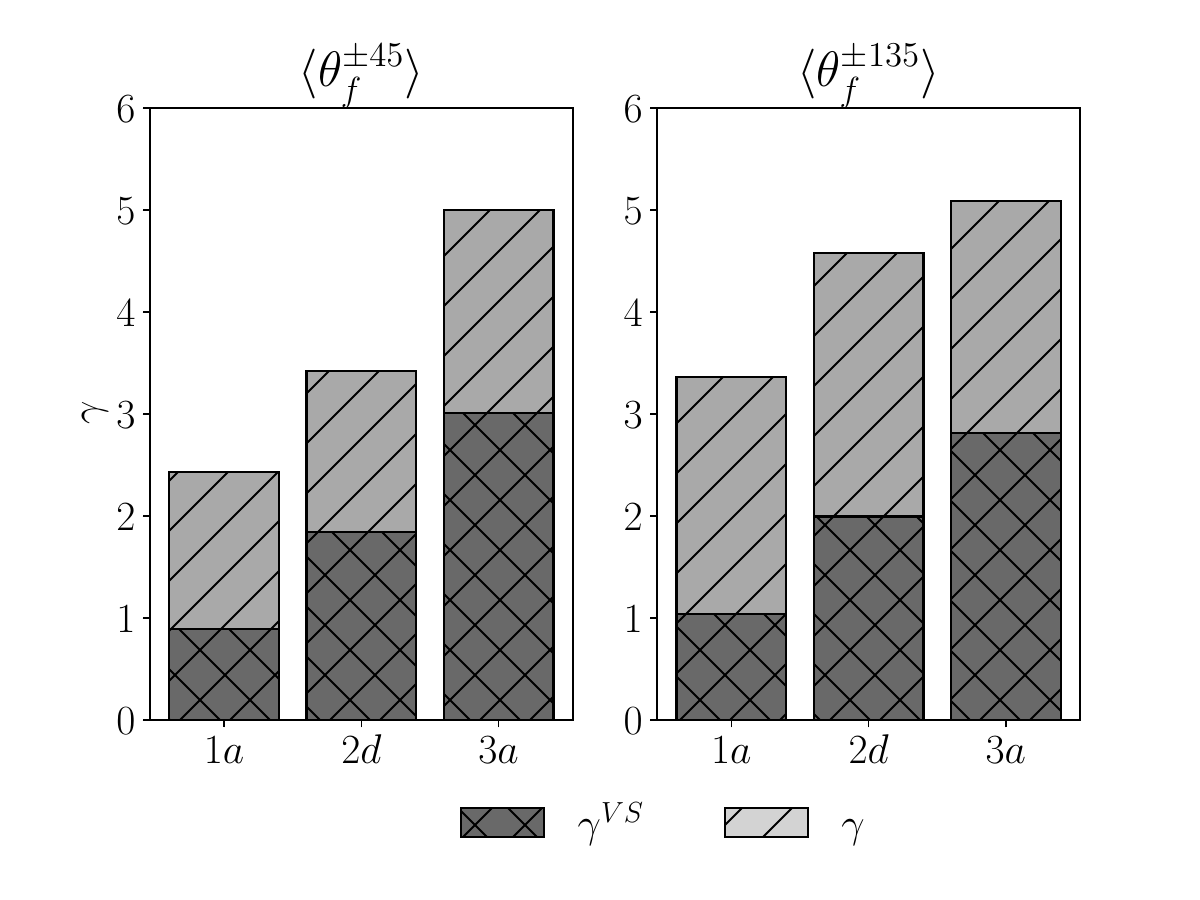}
	\raisebox{2in}{\textit{b})}\raisebox{-0.05in}{\includegraphics[width=.5\columnwidth]{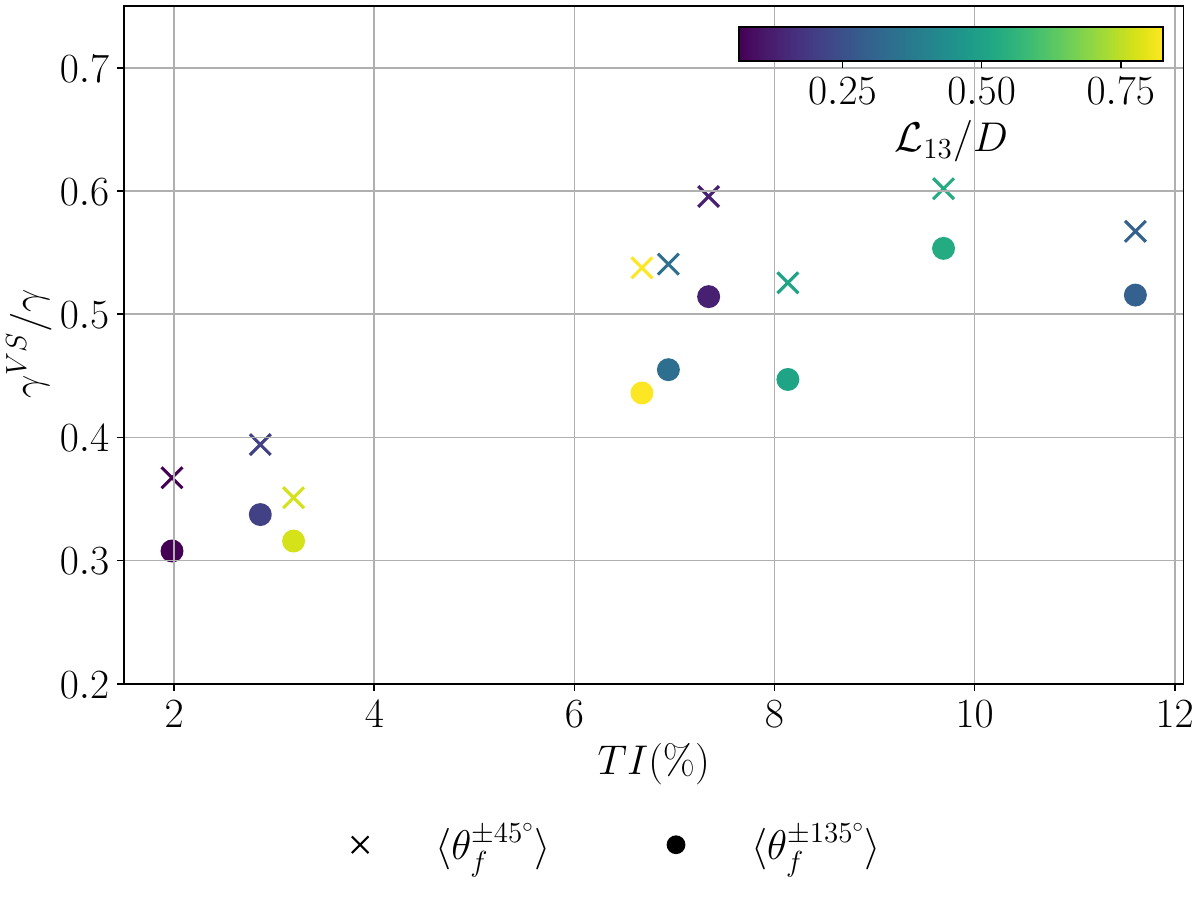}}}
	\caption{\textit{a}): characterisation of the contribution of regular vortex shedding to $\gamma$. \textit{b}): relative contribution of regular vortex shedding to $\gamma$ to the global induced root bending moment, for each FST case.}
	\label{fig:Figure7.3}
\end{figure}

Figure \ref{fig:Figure7.3} \textit{a)} presents $\gamma$ and $\gamma^{VS}$ for FST cases $1a$, $2d$, and $3a$ averaged between the windward and leeward faces ($\langle\theta^{\pm 45^{\circ}} \rangle$ and $\langle\theta^{\pm 135^{\circ}} \rangle$). The increase of $TI$ accounts for an increase of both $\gamma$ (as previously seen), and $\gamma^{VS}$. The increase of $\gamma^{VS}$ is in line with the increase of the energy associated with regular vortex shedding in the near wake of the cylinder, introduced by the presence of FST. Both the leeward and windward faces present similar magnitudes for $\gamma^{VS}$, suggesting that vortex shedding contributes mainly to the global deformation of the cylinder, inducing similar (but oppositely signed) strain dynamics in the leeward and windward faces. Figure \ref{fig:Figure7.3} \textit{b)} presents the evolution of the ratio $\gamma/\gamma^{VS}$ for each FST condition. The increase of $TI$ in FST accounts for a global increase in the contribution of the fluctuating root bending moment stresses in both faces, with the leeward face consistently presenting a smaller contribution from regular vortex shedding. This is likely to be introduced by the direct proximity between the several flow structures generated on wake that actively impact this surface. As seen from the autocorrelation evolution of the fluctuating strain on the windward face of the cylinder (see figure \ref{fig:Figure3.3}), this region mainly affected by the dynamics vortex shedding, in line with the larger $\gamma/\gamma^{VS}$.

\section{Conclusions}\label{sec:Conclusions}

The effect of FST on the flow dynamics, and on the structural response of a cantilevered cylinder with aspect ratio $\approx 10$, exposed to a turbulent cross-flow was assessed. To do so, a combination of $2$D-PIV and RBS sensors were used, capturing the flow velocity field and structural response concurrently. Thanks to the spatial information retrieved by the RBS network, we are able to assess the influence of the developing flow structures on the flow along the extent of the cylinder, before and after the shear-layer separation of the flow, allowing to assess the direct and indirect effect of FST. The cylinder was exposed to different conditions of FST, exploring different \enquote{flavours} of FST, to individually explore the effects of $\mathcal{L}_{13}/D$ and $TI$. Due to the $3$D nature of the flow under analysis, two FOVs were used at different spanwise locations to characterise the flow structures shed by the cylinder. The FOV focused on a region upstream, and downstream of the cylinder, capturing the free-stream and near-wake conditions to which the cylinder was subjected to concurrently to the distributed structural strain.   

The presence of FST was seen to affect the structural dynamics of the cylinder through three mechanisms: (1) reduced vortex formation length, (2) increased energy associated with wake flow structures, and (3) an enhanced spanwise coherence of these structures. Together, these factors intensify the cylinder's structural response. Specifically, FST increased the time-averaged loads by shifting the vortex formation region closer to the cylinder, thereby amplifying the root bending moment. These were further evidenced by increased strain fluctuation coherence along the cylinder's span, with the increase of $TI$ in FST.

Regular vortex shedding was seen to be energised especially in the near-wake, with the increase of $TI$ in FST, which mapped into a larger load magnitude associated with this flow structure. This increase was evident in the cross-power spectral densities (CPSD) between the velocity and strain fields, indicating a stronger correlation between flow fluctuations and the structural response under higher turbulence intensities. Our results also showed that FST tended to homogenize the strain dynamics across the cylinder's leeward and windward faces, suggesting that FST reduces spatial variations in response, and reinforces the dominant role of coherent flow structures like vortex shedding.

Through modal decomposition, we isolated the impact of vortex shedding, which was identified as the main contributor to the increased bending stresses under FST. This allowed us to quantify the impact of these coherent flow structures, assessing their relevance on the contribution to the global coupling of the structural and flow dynamics. Despite being the largest contributor for the respective structural dynamics, the near-wake correlation to the structural dynamics still shows a large impact of small-scales and other near-wake flow structures.

\textbf{Declaration of interests.} The authors report no conflict of interest.

\textbf{Availability of data and materials.} Data is available upon request to the corresponding author. For the purpose of open access, the authors have applied a Creative Commons Attribution (CC BY) licence to any Author Accepted Manuscript (AAM) version arising.

\appendix
\section{Mode clustering algorithm}\label{appendix}

Here we give a short and general overview of the algorithm developed in \citep{beit2021}. After employing OMD appealing to the modal expansion:
\begin{equation}
	x_j \approx \sum_{i=1}^{r}\alpha_i\phi_i^{*}e^{\lambda^{*}_i},
\end{equation}
where $\phi_i\in\mathcal{C}^{p}$ corresponds to the OMD modes with associated eigenvalue $\lambda^{*}_i$, and $\alpha_i\in\mathcal{C}$ are the coefficients associated with the modes obtained from the projection of $\phi_i$ onto the temporal velocity snapshots. We cluster the available modes with characteristic frequency $St_i$ taking into account their spectral and spatial similarity respectively defined as:

\begin{equation}
	\epsilon_{ij} = \mathrm{cos}(\mathrm{sin}^{-1}(\vert\vert \phi_i - \phi_j(\phi_i^T\phi_j)\vert\vert_2)), 
\end{equation}
\begin{equation}
	s_{ij} = \frac{1}{1+\vert St_i-St_j\vert^2}.
	\label{eq:spectral_filt}
\end{equation}

Having computed the spatial $\epsilon_{ij}$ and spectral $s_{ij}$ similarity matrices, we define the clustering matrix:
\begin{equation}
	d_{ij} =
	\begin{cases}
		1, & \text{if } \epsilon_{ij} - \delta_{ij} \geq \epsilon_0 \, \mathrm{and} \, s_{ij} \geq s_0, \\
		0, & \text{if } \epsilon_{ij} - \delta_{ij} < \epsilon_0 \, \mathrm{and} \, s_{ij} < s_0,
	\end{cases}
	\end{equation}
where $\epsilon_0$ and $s_{0}$ represent user defined cut-off parameters defined between $0\leq\epsilon_0\leq 1$ and $0\leq s_{0}\leq 1$. The final similarity matrix $d_{ij}$ is then iteratively computed by scanning through the available modes provided by OMD, grouping similar modes defined by $d_{ij}=1$. Given that the most energetic and coherent flow structure present is vortex shedding, the algorithm creates a single cluster $\mathcal{M}_{VS}$, defined by the set of modes selected $\mathcal{C}_{M} := \{\phi_{d_{ij}==1}\}$:
\begin{equation}
	\mathcal{M}_{VS} = SVD(\mathcal{C}_{M}\{\phi_{d_{ij}==1}\}).
\end{equation}

\begin{figure}
	\centering
	\raisebox{1.8in}{\textit{a})}\includegraphics[width=.45\columnwidth]{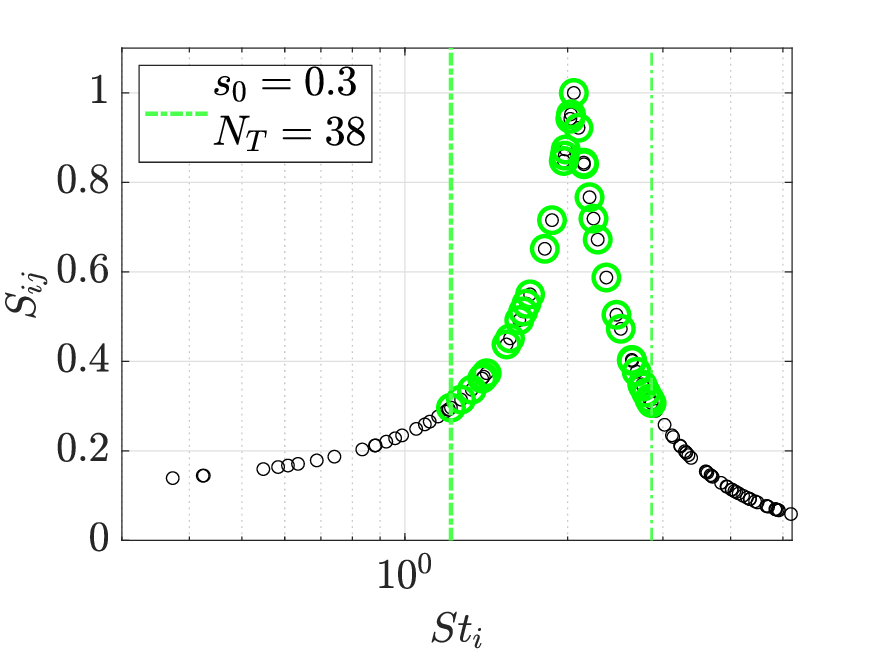}
	\raisebox{1.8in}{\textit{b})}\includegraphics[width=.45\columnwidth]{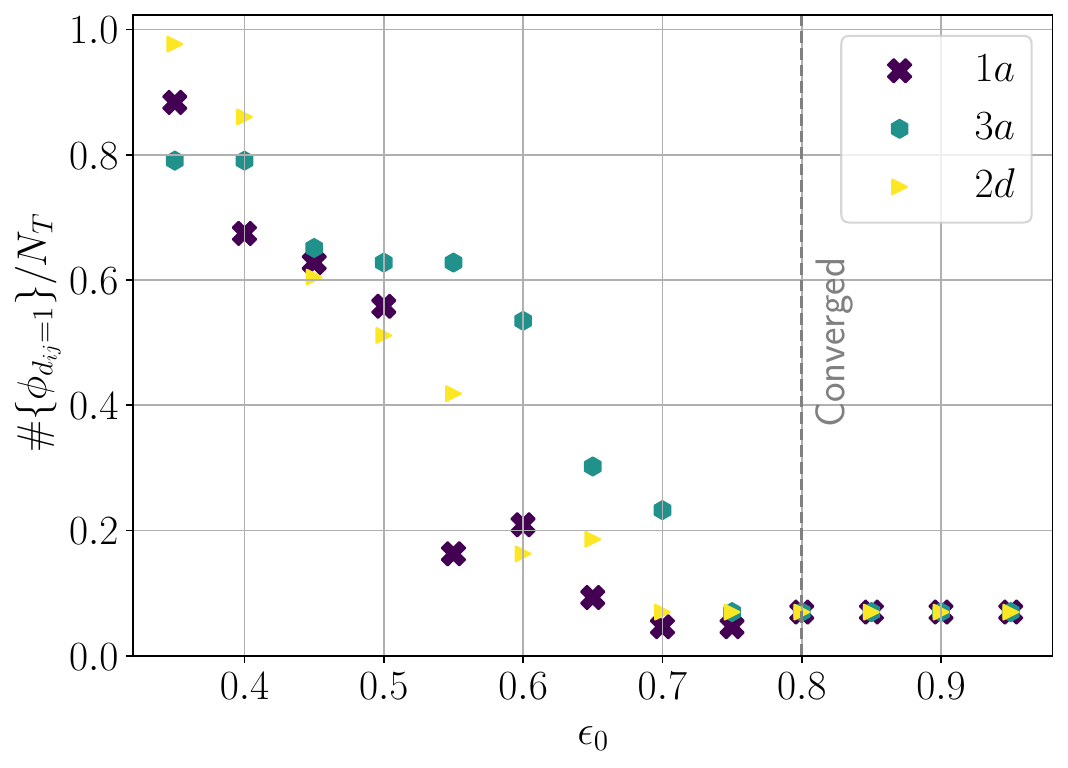}
	\caption{\textit{a)}: Selection of modes from spectral similarity around the vortex shedding. \textit{b)}: convergence of the algorithm with the increase of the spatial similarity cut-off parameter.}
	\label{fig:appendix}
\end{figure}

The spectral similarity cut-off parameter was set to be $s_{0} = 0.3$, ensuring that enough modes centred around vortex-shedding were assessed whether or not to be clustered. Figure \ref{fig:appendix} \textit{a)} presents the mode selection with equation \ref{eq:spectral_filt} for FST case $3a$ with $s_{0} = 0.3$, resulting in $N_T = 38$ modes to be assessed. The spatial similarity cut-off parameter was set to be $\epsilon_0 = 0.8$, minimizing the number of clustered modes selected and ensuring the algorithm’s convergence i.e., increasing the  cut-off value beyond this point did not affect the selected modes to be clustered - see figure \ref{fig:appendix} \textit{b)}. The decay rate and associated frequency of the final mode cluster are estimated from the clustered group of modes real and imaginary parts of their respective eigenvalues as follows:

\begin{equation}
	Im_{\mathcal{M}_{VS}} = \frac{1}{\#{\phi_{d_{ij}}==1}}\sum_{i\in \mathcal{C}_{M}\{\phi_{d_{ij}}==1\}} Im(\lambda_i^{*}),
\end{equation}
\begin{equation}
	Re_{\mathcal{M}_{VS}} = \left(\frac{1}{\#{\phi_{d_{ij}}==1}}\sum_{i\in \mathcal{C}_{M}\{\phi_{d_{ij}}==1\}} 1/Re(\lambda_i^{*})\right) ^{-1},
\end{equation}
where $\#{\phi_{d_{ij}}==1}$ corresponds to the number of modes forming the cluster $\mathcal{C}_{M}\{\phi_{d_{ij}}==1\}$.

\bibliographystyle{abbrvnat}

\end{document}